\documentclass[twoside,twocolumn]{article}

\usepackage[utf8]{inputenc}

\usepackage[sc]{mathpazo} 
\usepackage[T1]{fontenc} 
\usepackage{lmodern} 
\usepackage{microtype} 

\usepackage[english]{babel} 

\usepackage{graphicx}
\usepackage{color}
\usepackage[top=2cm,bottom=2cm,left=1.5cm,right=1.5cm,columnsep=20pt]{geometry} 
    {\large\scshape Computational Details%
    \par\medskip\normalfont\normalsize}%
    {}%
\newenvironment{acknowledgement}
    {\large\scshape Acknowledgement%
    \par\medskip\normalfont\normalsize}%
    {}%

    {\large\scshape Supporting Information%
    \par\medskip\normalfont\normalsize}%
    {}%
\usepackage{multirow} 
\usepackage{mathtools} 
\usepackage[normal, small,labelfont=bf,up,textfont=it,up]{caption} 
\usepackage{booktabs} 

\usepackage{lettrine} 

\usepackage{enumitem} 
\setlist[itemize]{noitemsep} 

\usepackage{abstract} 

\usepackage{titlesec} 
\renewcommand\thesection{\Roman{section}} 
\renewcommand\thesubsection{\roman{subsection}} 
\titleformat{\section}[block]{\large\scshape\centering}{\thesection.}{1em}{} 
\titleformat{\subsection}[block]{\large}{\thesubsection.}{1em}{} 

\usepackage{fancyhdr} 
\usepackage{titling} 

\usepackage{hyperref} 

\pretitle{\begin{center}\Large\bfseries} 
\posttitle{\end{center}} 

\def\bea{\begin{eqnarray}}
\def\eea{\end{eqnarray}}
\def\ben{\begin{equation}}
\def\een{\end{equation}}
\def\benu{\begin{enumerate}}
\def\enu{\end{enumerate}}

\def\bei{\begin{itemize}}
\def\eei{\end{itemize}}
\def\beit{\begin{itemize}}
\def\eit{\end{itemize}}
\def\benu{\begin{enumerate}}
\def\enu{\end{enumerate}}

\def\n{n}

\def\sss{\scriptscriptstyle\rm}

\def\1var{(\bx_1...\bx\N)}

\def\br{{\bf r}}

\def\bx{{x}}

\def\s{_{\sss S}}
\def\xc{_{\sss XC}}

\def\N{_{\sss N}}

\def\HF{^{\rm HF}}

\def\LDA{^{\rm LDA}}

\def\sph_int{ {\int d^3 r}}

\def\intr{\int d^3r\,}

\usepackage[table]{xcolor}
\usepackage{amsmath}
\usepackage{graphicx}

\usepackage[]{quoting}
\usepackage{multirow} 

\newcommand\ssh{\textcolor{magenta}} 

\def\d{_{\sss D}}
\def\f{_{\sss F}}

\title{Extending density functional theory with
near chemical accuracy beyond pure water}

\author{%
\textsc{Suhwan Song$^{a,b}$, Stefan Vuckovic$^{c,d}$, Youngsam Kim$^a$, Hayoung Yu$^a$,} \\ \textsc{Eunji Sim$^{a,}$\thanks{esim@yonsei.ac.kr}, and Kieron Burke$^{b,e}$} \\ 
\normalsize $^a$Department of Chemistry, Yonsei University, 50 Yonsei-ro Seodaemun-gu, Seoul 03722, Korea \\ 
\normalsize $^b$Department of Chemistry, University of California, Irvine, CA 92697, USA \\  
\normalsize $^c$Institute for Microelectronics and Microsystems (CNR-IMM), Via Monteroni,Campus Unisalento, 73100 Lecce, Italy \\
\normalsize $^d$Department of Chemistry\&Pharmaceutical Sciences and Amsterdam Institute of Molecular and Life Sciences (AIMMS), \\
\normalsize Faculty of Science, Vrije Universiteit, De Boelelaan 1083, 1081HV Amsterdam, The Netherlands\\
\normalsize $^e$Departments of Physics \& Astronomy, University of California, Irvine, CA 92697, USA \\  
}
\date{\today} 

\usepackage{xr}
\usepackage{float}

\renewcommand{\maketitlehookd}{%
\begin{abstract}
\noindent 
Density functional simulations of condensed phase water are typically inaccurate, due to the inaccuracies of approximate functionals. A recent breakthrough showed that the SCAN approximation can yield chemical accuracy for pure water in all its phases, but only when its density is corrected.  
This is a crucial step toward first-principles biosimulations.  
However, weak dispersion forces are ubiquitous and play a key role in noncovalent interactions among biomolecules, 
but are not included in the new approach.  
Moreover, na\"ive inclusion of dispersion in HF-SCAN ruins its high accuracy for pure water. 
Systematic application of the principles of density-corrected DFT 
yields a functional (HF-r$^2$SCAN-DC4) which recovers and not only 
improves over HF-SCAN for pure water, 
but also captures vital noncovalent interactions in biomolecules, making it suitable for simulations of solutions.
\end{abstract}
}

\begin{document}

\maketitle
\sf
\small

\section{Introduction}
\label{sec_intro}


{\bf The overwhelming importance of simulating water:}
The properties of water, such as the uniqueness of its phase diagram,
never stop surprising scientific communities.\cite{ZWCW21}
Given the vital importance of water in fields that vary from material science to biology,
there has been a recent surge in the development and competition of different electronic structure
methods for simulating water.\cite{SWSA20,DLPP21,DSBP22,ZCTX21,TPFR21,LHP21,PLSH22,PLDP22} 
As {\em ab initio} quantum-chemical methods are too expensive for large systems,
Kohn-Sham density functional theory (KS-DFT) has become a workhorse 
of electronic structure methods 
for running water calculations.\cite{BLP13, BMP14, MBP14,RSBP16,CKRC17}
But, despite an excellent accuracy to cost ratio, historically KS-DFT has been unable 
to deliver sufficiently high accuracy in water simulations 
to reproduce experimental data.\cite{KMMS04,KKP09,SKAT11,GAM16}

\begin{figure*}[htb]
\centering
\includegraphics[width=1.95\columnwidth]{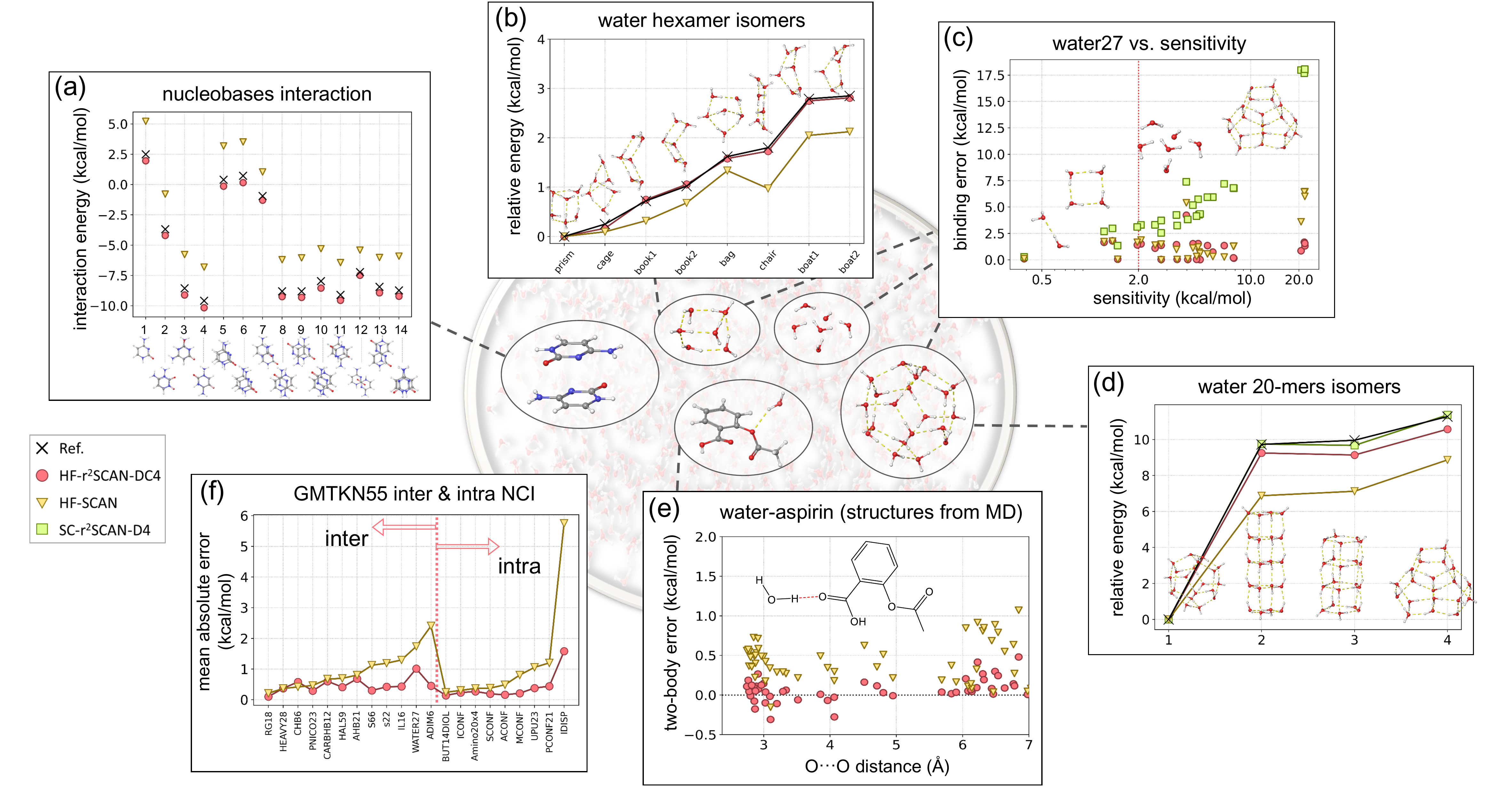}
\caption{
Performance of HF-r$^2$SCAN-DC4 relative to HF-SCAN
for various chemical reactions:
(a) the interaction energy of various configurations of
the stacked cytosine dimer, where HF-SCAN underbinds by 2-3 kcal/mol;
(b) energies of water hexamer relative to the lowest-lying 
prism isomer, with HF-SCAN underestimating by almost 1 kcal/mol;
(c) errors in binding energy of WATER27 complexes as a
function of density sensitivity (how much a DFT energy changes when
the density is changed), showing how large errors 
can be without
using
the HF density. 
One cluster,  
H$_3$O$^+$(H$_2$O)$_6$OH$^-$ (at $x$ close to 4 kcal/mol)
is an outlier argued to  exhibit a significant 
multiconfigurational character\cite{DSBP22};
(d) relative energies of water 20-mer isomers (not density sensitive)
from WATER27, where self-consistent SC-r$^2$SCAN-D4 performs best, but
using the HF density introduces little error;
(e) errors in interaction energies in the 
water$\cdots$aspirin dimer structures from an MD simulation at T=298.15 K;
(f) mean-absolute-errors (MAEs) for intra- and inter-molecular
noncovalent interactions datasets from the GMTKN55 database.
For more details, see the main text and supporting information.}
\label{fgr:1}
\end{figure*}

A recent breakthrough in this direction by Dasgupta {\it et al.} showed
that the strongly constrained and appropriately normed (SCAN) functional,
when used in tandem with `density-corrected DFT' (DC-DFT),
is a game changer for water simulation, because it brings
KS-DFT close to chemical accuracy.\cite{DLPP21,DSBP22}
The role of water in a chemical or biochemical reaction goes beyond providing an environment to help a  reaction in an aqueous solvation and is often explicitly involved in the mechanism.  
For this reason, a complete understanding of the reaction is possible only when the interaction between water and other molecules is accurately described.
Figure~\ref{fgr:1} shows how an integratively designed DC-DFT procedure,  HF-r$^2$SCAN-DC4,  describes not only the interactions between water-water, water-organic molecules, and water-biochemical molecules in various situations, but also the interactions of noncovalent complexes at chemical accuracy or better.

{\bf The importance of the density:}
DC-DFT is a general framework that separates errors of any DFT calculations into
a contribution coming from the approximate "D" (density) and the 'true' error coming from the approximate "F" (functional).\cite{KSB13,KSB14,WNJK17,VSKS19}
In addition to being a rigorous exact theory,  DC-DFT gives practical guidance on when and how it
can be used to reduce errors in DFT simulation.\cite{SSB18,NSSB20,JPSC20,SVSB21}
Standard DFT calculations are performed self-consistently (SC).
The simplest form of practical DC-DFT is HF-DFT, where density functionals are evaluated instead
on Hartree-Fock (HF) densities and orbitals.\cite{KSB11,KPSS15,SKSB18,KSSB18,SVSB22}
While in most cases, SC-DFT gives the best answer,  in some errors in
specific cases SC-DFT suffers from large energetic errors due to the approximate density (density-driven errors).\cite{KSB13,VSKS19}
In such cases,  HF-DFT typically yields significant improvements over SC-DFT,  
and these include a number of chemical domains (barrier heights,  some torsional barriers,  halogen bonds,  anions,  etc.).\cite{SVSB21}

{\bf The importance of the functional:}
SCAN is a non-empirical
meta-GGA functional designed to satisfy 17
exact physical constraints, and to recover several nonbonded `norms'.\cite{SRP15}
Meta-GGA's use the KS kinetic energy density as an ingredient, but are {\em not} hybrid functionals like B3LYP\cite{B88,LYP88,B93,SDC94},
which include some fraction of exact exchange from a HF calculation.\cite{B93}
In terms of accuracy,  SCAN is often on par with highly empirical more expensive density functionals designed for molecules.
At the same time, it enjoys great successes for simulations of extended systems,
making it one of the most-used general-purpose functionals developed 
over the last 10 years.\cite{SRZS16,TSB16,MH17,GHBE17,ZSPW17}
Earlier works have shown that

\noindent standard (SC) DFT calculations of water clusters suffer badly
from density-driven errors,
which explains why HF-SCAN is much more accurate than its SC counterpart for simulations of water.\cite{DLPP21,JPSC20}
In addition to water clusters,  Dasgupta {\it et al.} used HF-SCAN in tandem with many-body potential
energy function related to the highly popular MB-pol\cite{BLP13,BMP14,MBP14} to run molecular dynamics (MD) simulations of liquid water
and obtained results in excellent agreement with the experimental data.  
These were the first successful DFT-based simulations able to correctly describe the condensation of water.

Nevertheless, the convergence of the SCAN functional can be painfully slow with respect to the size of molecular grids,
due to either the size of a system
or because it would require grids larger than those available in most of the standard-quantum chemical codes.\cite{SVSB22}
Larger grids also lead to longer computational times.  
To address these issues of SCAN, Perdew and co-workers developed the regularized-restored SCAN functional (r$^2$SCAN), 
which regularizes SCAN but restores SCAN’s adherence to exact constraints.\cite{FKNP20}
But, as we show below, a standalone version of HF-r$^2$SCAN is much less accurate for water simulations than HF-SCAN.

{\bf The vital importance of dispersion:}
Despite enormous success in modelling water,  HF-SCAN is not a panacea.  
In their water simulations,  Dasgupta {\it et al.} used HF-SCAN without dispersion correction,
 as they found that the standard dispersion corrections,  such as those of Grimme\cite{GAEK10},
{\em  worsen} the original results of HF-SCAN for water.
But such dispersion corrections have long been known to be necessary 
for noncovalent interactions (NCIs).\cite{WVNL01,MS96,G04,BJ05,G06a,BJ07,TS09,VV10,RH16,PBJ21}
So, despite delivering a high accuracy for pure water simulations,  
HF-SCAN without a dispersion correction cannot describe accurately long-range dispersion interactions. 
For this reason,  the errors of HF-SCAN are several times larger than
those of DFT enhanced by a dispersion correction for the standard noncovalent datasets.\cite{GHBE17}
The challenge is then to construct an efficient density functional that
correctly describes NCIs of different nature, 
while recovering or even improving the accuracy of HF-SCAN for water simulations.

{\bf HF-r$^2$SCAN-DC4, an integratively designed DC-DFT procedure:}
In the present paper,  we resolve these issues by using
the principles of DC-DFT to carefully parameterize a dispersion correction for HF-r$^2$SCAN.  
This yields
HF-r$^2$SCAN-DC4,  which 
produces the following key results:
(i) HF-r$^2$SCAN-DC4 {\em improves} upon HF-SCAN for pure water simulations,  
by up to 0.7 kcal/mol for relative energies of water hexamers,
 and up to 2.4 kcal/mol for those of water 20-mers;
(ii) HF-r$^2$SCAN-DC4 is far more accurate than HF-SCAN for interactions of water with other
molecules and for NCIs in general, because of the inclusion of explicit dispersion
corrections;
(iii) HF-r$^2$SCAN-DC4 can be routinely and efficiently used in calculations because,
unlike HF-SCAN\cite{SVSB22},  HF-r$^2$SCAN-DC4 has no grid convergence issues.
In our HF-r$^2$SCAN-DC4,  each of the three ingredients is vitally important:
The "HF" part reduces density-driven errors, while
r$^2$SCAN fixes the grid issues of SCAN.
But most importantly, the way in which we parametrize the D4 corrections
by using the DC-DFT principles is vital, as an unwitting fitting of D4 
ruins the accuracy for water simulations.
If we drop {\bf any} of those elements of
HF-r$^2$SCAN-DC4,  at least one of its three appealing results will be lost.

\begin{figure*}[h]
\centering
\includegraphics[width=1.95\columnwidth]{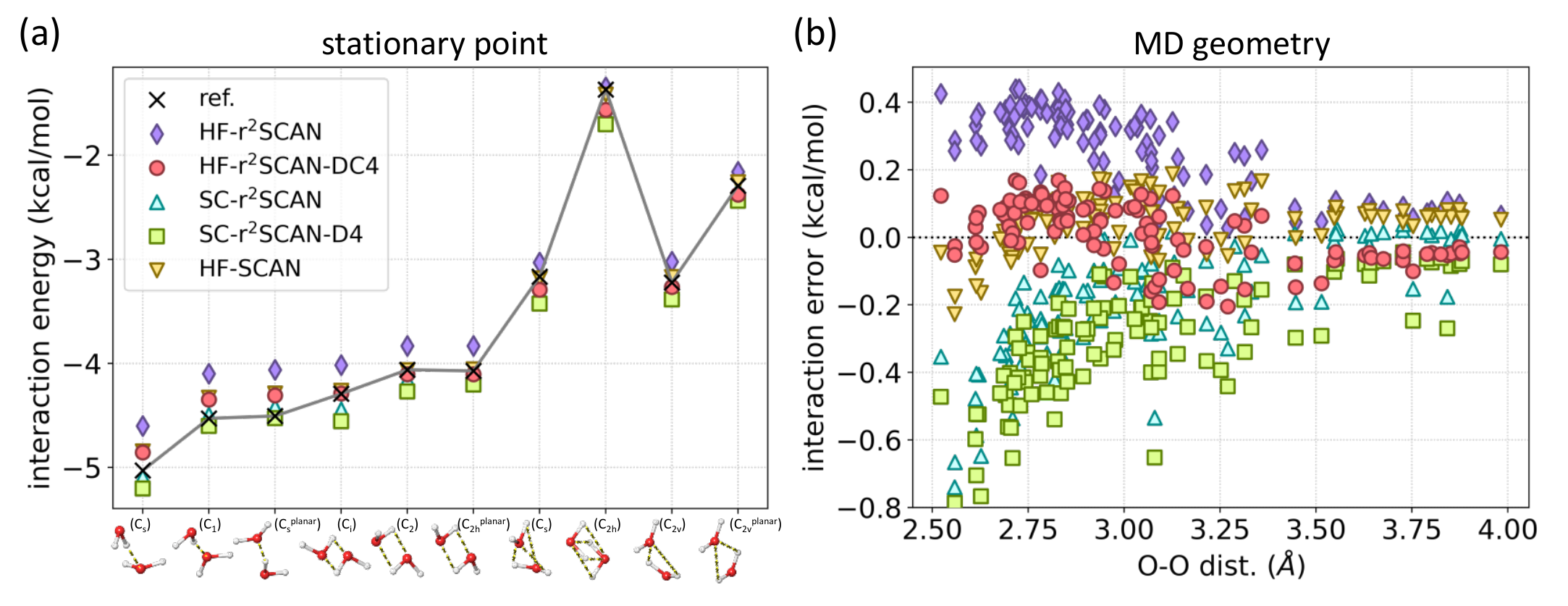}
\caption{
Water dimer interaction energies for 
(a) Smith stationary points\cite{SSPS90} and 
(b) MD simulated water dimers with the oxygen-oxygen distance.
For (a), 
MAEs of each functional are (following the order in the legend)
0.25, 0.11, 0.09, 0.17, and 0.08 kcal/mol. 
DLPNO-CCSD(T)-F12 has been used as a reference.
For (b),
MAE of each functionals are
0.25, 0.08, 0.20, 0.30 and 0.08 kcal/mol.
Figure~\ref{fgr:dimer_s} shows the corresponding density sensitivities 
and Figure~\ref{fgr:dimer_err} shows the errors of approximations for the Smith dimers and 
interaction energies for MD dimers.
}
\label{fgr:dimer}
\end{figure*}

To illustrate all these points, and how they work together, we created
Figure~\ref{fgr:1}.  We show how HF-r$^2$SCAN-DC4 is better than HF-SCAN for interactions of 
nucleobases [panel (a)], 
water molecules with one another [panels (b),  (c),  (d)],  
water with other molecules [panel (e)],  and NCIs in general [panel (f)].

Stacking interactions in nucleobases are of vital importance in biology as their energetics
is essential to describe the formation and stability of DNA and RNA.\cite{YT05,  KS19} 
In Figure~\ref{fgr:1}(a), we compare the accuracy of 
HF-SCAN and HF-r$^2$SCAN-DC4 for interaction energies 
of stacked cytosine dimers at different configurations.
As we can see from Figure~\ref{fgr:1}(a),
our HF-r$^2$SCAN-DC4 essentially 
greatly reduces the errors of 
HF-SCAN that systematically underbinds these stacked complexes by about 2.5 kcal/mol. 
This demonstrates that despite its success for modeling water, 
HF-SCAN misses most of dispersion and thus cannot compete with our HF-r$^2$SCAN-DC4 in
modelling NCIs.  This is especially the case
for NCIs dominated by dispersion interactions as those present in stacked nucleobases.  
(See Figure~\ref{fgr:ks19_err} 
in the supporting information
for the errors in interaction energies.)
We note that the mean absolute error (MAE) of 
HF-r$^2$SCAN-DC4 (0.4 kcal/mol) is very good relative to HF-SCAN,
but not very impressive relative to B3LYP-D3(BJ) (less than 0.2 kcal/mol).\cite{KS19}  
But such functionals
include only a fraction of HF exchange, and so still suffer from large density-driven errors in water,
and so have larger errors for pure water (as shown below).

\begin{figure*}[htb]
\centering
\includegraphics[width=1.95\columnwidth]{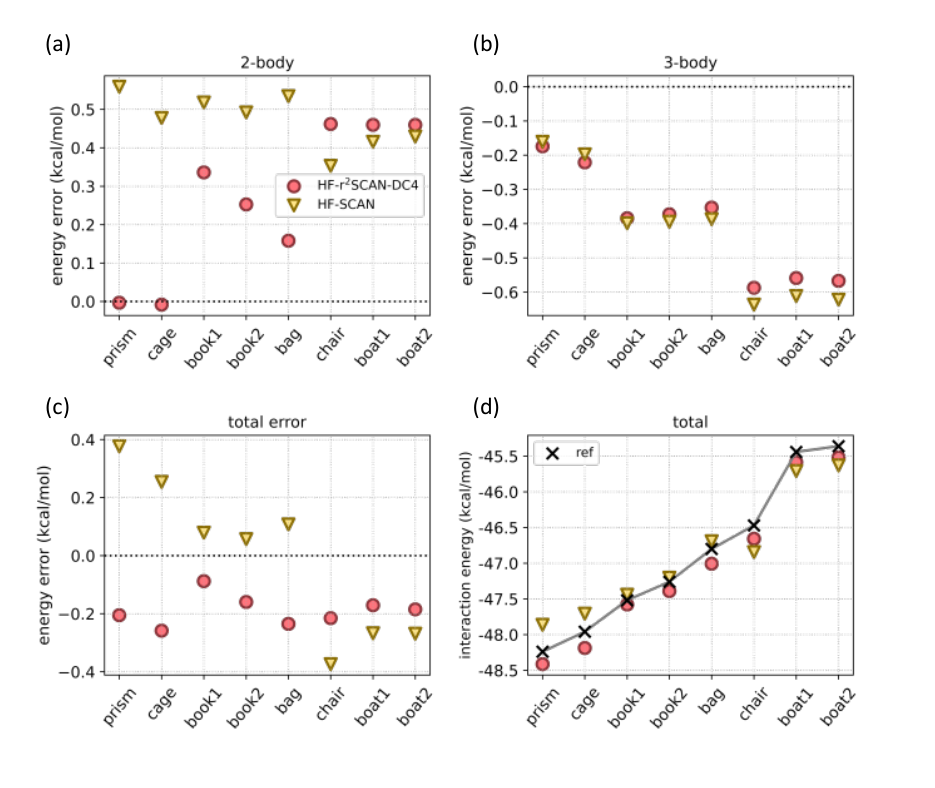}
\caption{
K-body interaction energy errors with (a) $K=2$, (b) $K=3$,
and (c) total, and (d) the interaction energy
for 8 water hexamers.
(For higher order $K$-body interaction energies, see 
Figure~\ref{fgr:nbodyerr} and~\ref{fgr:nbodypor}.)
Geometries and
CCSD(T)/CBS reference interaction energies are from Ref.~\cite{RSBP16}.
The MAEs of HF-r$^2$SCAN-DC4 and HF-SCAN are
0.19 kcal/mol and 0.22 kcal/mol,  respectively.
}
\label{fgr:nbody}
\end{figure*}


Water hexamers, "the smallest drops of water"\cite{NM00,WBBP12},   
are important, 
as they represent the transition from 
two-dimensional to three-dimensional 
hydrogen-bonding networks.\cite{BT09,CL10,OJ13} 
The energy differences between two adjacent isomers of water hexamers are tiny,
making even the ordering of isomers a very challenging test for quantum-chemical methods.\cite{BT09,SMFT08}
In Figure~\ref{fgr:1}(b),  we compare the energies of water
hexamer isomers relative to the energy of the prism,  as the lowest-lying isomer.\cite{BT09,OBKS07,GMTA12}
Despite being more accurate 
for water hexamers than
most 
DFT methods available on the market,  
HF-SCAN mistakes the ordering of the isomers, as it predicts too low energies of the chair isomer.
Our HF-r$^2$SCAN-DC4 
is also here superior to HF-SCAN,
as it not only gives
the right ordering of isomers,  
but essentially reproduces the reference values for the relative energies of isomers.
If D4 is fitted by not accounting for the DC-DFT principles (see below), 
the accuracy of HF-r$^2$SCAN-DC4 for the water simulation is lost.
This happened in Ref.~\cite{SM21b} and will be discussed in the method section.

We use the WATER27 dataset to illustrate the importance (and subtlety) of
DC-DFT for water simulations.
WATER27 is a standard 
dataset for binding energies of water clusters.
Density sensitivity,
$\tilde S$,  
is a measure for how sensitive a given DFT simulation is to errors in densities (see Section~\ref{sec:dde}
in the supporting information for further details and specific definitions).\cite{SSB18}
Typically,  the errors of SC-DFT calculations grow with $\tilde S$,
indicating the presence of large density-driven errors.\cite{SVSB21,SVSB22,SSVB22}
DC-DFT reduces these large density-driven errors of SC-DFT and thus 
the errors of DC-DFT do not grow with $\tilde S$. 
In Figure~\ref{fgr:1}(c) we plot WATER27 errors as a function of density sensitivity.
As
the errors of SC-r$^2$SCAN-D4 grow with $\tilde S$, so also does
the energetic improvement of HF-r$^2$SCAN-DC4 over SC-r$^2$SCAN-D4.
Furthermore,  
sometimes dispersion corrections worsen SC-DFT for cases with large density-driven errors.\cite{JPSC20,SVSB21}
This is also the case here, as 
SC-r$^2$SCAN-D4 significantly deteriorates the accuracy of SC-r$^2$SCAN
(see Figure  \ref{fgr:sc_water27}).
The errors of HF-SCAN are also substantially lower than those of SC-r$^2$SCAN-D4,
and for most of the binding energies of the WATER27 clusters, HF-SCAN is comparable to HF-r$^2$SCAN-DC4.  
But, for the four clusters with the largest sensitivities, 
HF-r$^2$SCAN-DC4 outperforms HF-SCAN by $\sim$4 kcal/mol.

WATER27 is a part of 
the GMTKN55\cite{GHBE17}, a database that we use to train the D4 parameters in HF-r$^2$SCAN-DC4 (see methods).
But,  according to the principles of DC-DFT,
we exclude those WATER27 clusters that are density-sensitive, as their energetic errors are dominated by
the errors in their densities.\cite{SVSB21}
Thus none of the clusters
that are to the right of the vertical dashed line placed at $\tilde S$ =2 kcal/mol (see Method Section 
for the details on this reasoning) are used in the fitting, which means
HF-r$^2$SCAN-DC4 makes genuinely accurate predications for a vast majority of these water clusters.
Not only does it recover
HF-SCAN for binding energies of the water clusters,
but also provides substantial improvements for the most challenging clusters. 

An important question is whether or not one should {\em always} correct the density.
The general principles of DC-DFT say that one should only correct the density in cases
of substantial density-driven errors.  In density insensitive cases, the effect of correcting
the density should be small, and may actually worsen energetics.
Figure \ref{fgr:1}(d) shows energies of water 20-mers relative to the energy of the lowest of the four 20-mers.
Here SC-r$^2$SCAN-D4 beats its DC counterpart every time.
In contrast to large $\tilde{S}$ for binding energies of the four 20-mers (the last four datapoints in Figure  \ref{fgr:1}(c)),
the sensitivities corresponding to their relative isomer energies are about twenty times smaller (see Figure  \ref{fgr:20mer_s}). 
Thus the higher accuracy of SC-r$^2$SCAN-D4 over HF-r$^2$SCAN-DC4 does not come as a surprise.
But the crucial point is that,
even in this {\em low-sensitivity} scenario, the errors introduced by the HF density are far smaller
than those of HF-SCAN, and remain tiny on a per molecule basis.

A crucial figure of merit is how accurate energetics are for water molecules in the vicinity of
an organic molecule, especially if it is polar.
In Figure \ref{fgr:1}(e),  we show errors in the interaction energies between water and
aspirin from structures that we extracted from an MD simulation at T=298.15 K
(see Section~\ref{sec:water} for further details on the MD simulation).
The structures are sorted by the distance between the oxygen atom in water 
and the specified oxygen atom in the carboxyl group of aspirin. 
The errors of HF-r$^2$SCAN-DC4 are much smaller than 
those of HF-SCAN.  They are also substantially smaller than those 
of SC-r$^2$SCAN-D4 (see Figure~\ref{fgr:aspirin_tmp}), 
demonstrating again the importance of both the D4 and DC components in our method.

Getting NCI right across a broad range of molecules is important,
even in the absence of water.   The GMTKN55 collection of 55 databases has become a
standard benchmark\cite{GHBE17} and includes many databases for NCIs.
In Figure~\ref{fgr:1}(f),  we 
compare the MAEs of HF-SCAN and HF-r$^2$SCAN-DC4 
for the standard datasets 
with intra- and intermolecular NCIs\cite{GHBE17}.
Despite its high accuracy for water clusters, HF-SCAN does not capture long-ranged dispersion interactions.
This is why it is far less accurate than HF-r$^2$SCAN-DC4 for noncovalent datasets. 
We can see that HF-r$^2$SCAN-DC4 is highly accurate here,  and 
on average it beats SC-r$^2$SCAN-D4 for both inter- and intramolecular NCIs
(see Table~\ref{tbl:gmtkn} in the supporting information comparing the metrics for overall performance).

\section{Results}

\subsection{Interaction energies for water dimers}

As discussed already, HF-SCAN performs incredibly well for interactions in pure water.   
In this section, we look at select water dimers that are relevant to water simulations,
and show how HF-r$^2$SCAN-DC4 reproduces (or even exceeds) this accuracy.
More importantly, we show how each aspect of its construction (density correction, regularization of SCAN, and
dispersion correction) are vital to its accuracy for water.  Later we will show that no other approximation
at this level of cost comes close to this performance for water.

Figure~\ref{fgr:dimer} shows the interaction energies for many water dimers
(the difference in the energies of a dimer and two monomers).
(a) shows the interaction energies at Smith stationary points, 
some of which resemble geometries from dense ice structures.\cite{GAM16}
(b) shows the errors of approximations in interaction energies for water
dimers as a function of the distance between the two oxygen atoms.  The underlying structures were
extracted from an MD simulation at T=298.15K (see Section~\ref{sec:water} for further details on the simulation). 
For the interaction energies of 
these water dimers,  HF-SCAN without a dispersion correction already 
provides a very high accuracy (with MAEs of less than 0.1 kcal/mol).
Our HF-r$^2$SCAN-DC4 essentially recovers this high accuracy of HF-SCAN.
Similar patterns observed for binding energies of water clusters are also seen here.

By studying the various plots, one can assess the importance of the relative contributions to
HF-r$^2$SCAN-DC4.  First, the purple points give HF-r$^2$SCAN, to be contrasted with HF-SCAN.
We see that HF-r$^2$SCAN significantly (on this scale) underestimates the interaction energy.
Even though r$^2$SCAN was designed to reproduce the results of SCAN, these differences are
so small as to be negligble for most purposes.   However, they are clearly significant here,
showing HF-r$^2$SCAN is noticeably less accurate for these dimers.  The addition of the D4
correction, however, makes their errors comparable.

On the other hand, we may also consider the importance of density correction.   We see that
SC-r$^2$SCAN-D4 considerably overestimates interaction energies.  In fact, SC-r$^2$SCAN
does rather well, as the errors due to poor density and missing dispersion cancel.

We can also observe from Figure~\ref{fgr:dimer}(b) that the improvement of 
HF-r$^2$SCAN-DC4 over SC-r$^2$SCAN-D4 decreases with the 
distance between the two oxygen atoms in water dimers.  
This can be understood in terms of 
underlying density sensitivity  which 
also decreases with the O-O distance (see Figure~\ref{fgr:dimer_s}).


\subsection{Many-body interactions in larger water clusters}

In Figure~\ref{fgr:nbody} we compare
errors of HF-r$^2$SCAN-DC4 and HF-SCAN
for the 
interaction energies of the eight standard water hexamers.\cite{BT09,CL10}
In addition to total interaction energies, 
we also use the 
many-body expansion (MBE) to
show the $K$-body contributions to these energies (with $K$ in between 2 and 6). 
This is a standard methodology for understanding the origins of errors 
in water models.\cite{DLPP21,RSBP16,GPCS11}
The energetic importance of the $K$-body contributions decreases rapidly
with $K$ (Figure~\ref{fgr:nbodypor}), 
making the 2-body contributions by far the most important, and these
are where significant differences emerge when the density is corrected.
But in order to reach chemical accuracy,
a proper description of the higher-order contributions also matters. 
The 2-body plot shows that HF-SCAN has a rather systemative overestimate of
about 0.5 kcal/mol, whereas
HF-r$^2$SCAN-DC4 is substantially less for about half the clusters.
The 3-body plot shows them being almost identical.  But 
in the total error, we see that HF-r$^2$SCAN-DC4 is far more systematic,
as HF-SCAN makes errors of opposite sign, while HF-r$^2$SCAN-DC4 
is always an overestimate of about 0.2 kcal/mol.   

This consistency is important on the plot (d), showing the interaction
energy of the 8 hexamers.   
Because HF-r$^2$SCAN-DC4 is so consistent, it
gets the ordering in interaction energies
of all clusters correct, whereas HF-SCAN incorrectly
predicts that the interaction energy in the 
bag is higher than that of the chair.
The MAE of HF-r$^2$SCAN-DC4 is 0.19 kcal/mol, 
lower than 0.22 kcal/mol
for HF-SCAN.
On average, HF-r$^2$SCAN-DC4
also improves individual $K$-body contributions to the interaction energies, except for $K=4$,
where
both are marginally small (Figure~\ref{fgr:nbodyerr}). 
This MBE test shows us that the improvement of 
HF-r$^2$SCAN-DC4 over HF-SCAN for the water hexamer interaction energies
(seen also for the relative isomer energies (Figure~\ref{fgr:1}(b)) is systematic
and does not result from the error cancellations between different $K$-body contributions
(for the detailed information of water hexamer isomerization energy in Figure~\ref{fgr:1}, 
see Figure~\ref{fgr:hex_rank}).

\subsection{Water$\cdots$cytosine interaction energies}

\begin{figure}[htb]
\centering
\includegraphics[width=0.95\columnwidth]{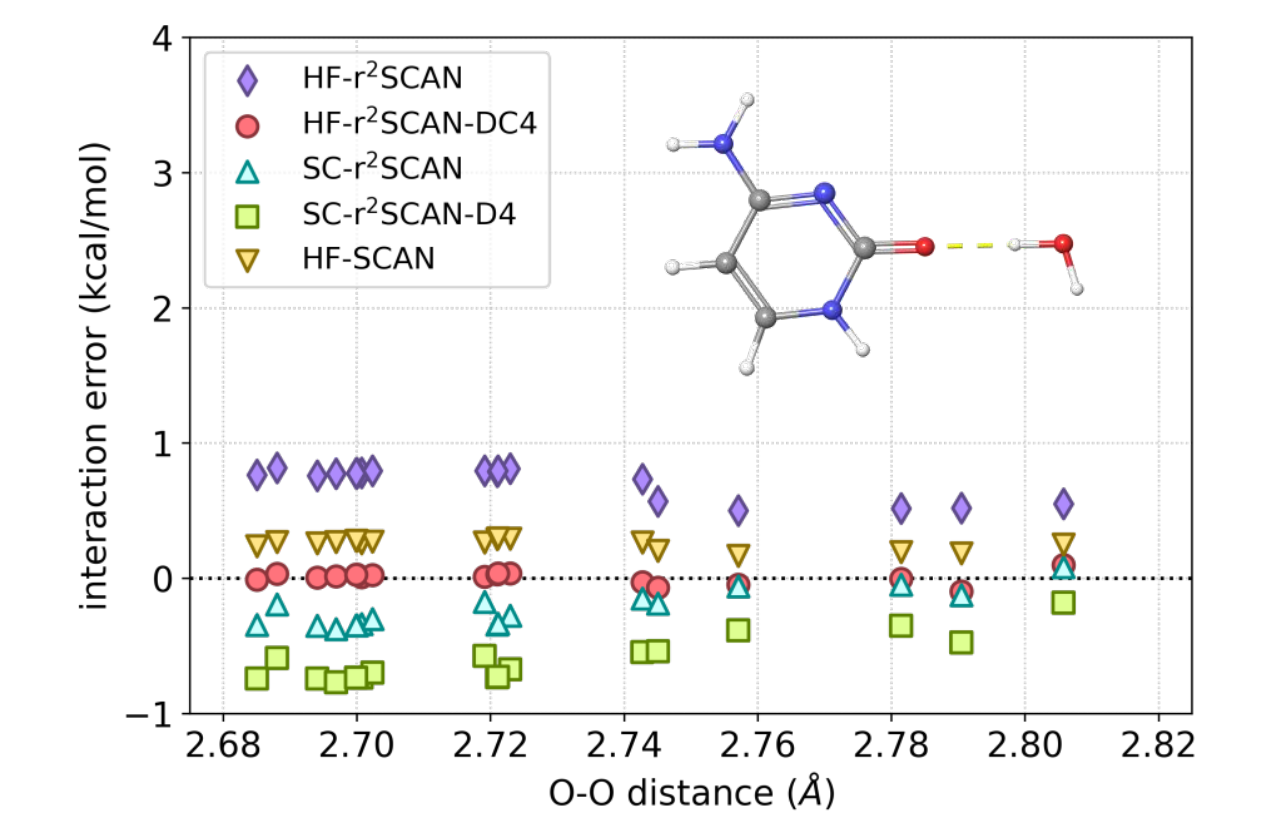}
\caption{
Errors in interaction energies of water$\cdots$cytosine complexes
sorted by the distance between the oxygen atom in cytosine and 
the oxygen atom in water.  
Reference interaction energies have been computed at the 
DLPNO-CCSD(T)-F12/AVQZ level of theory.
}
\label{fgr:wc}
\end{figure}

\begin{figure*}[htb]
\includegraphics[width=1.95\columnwidth]{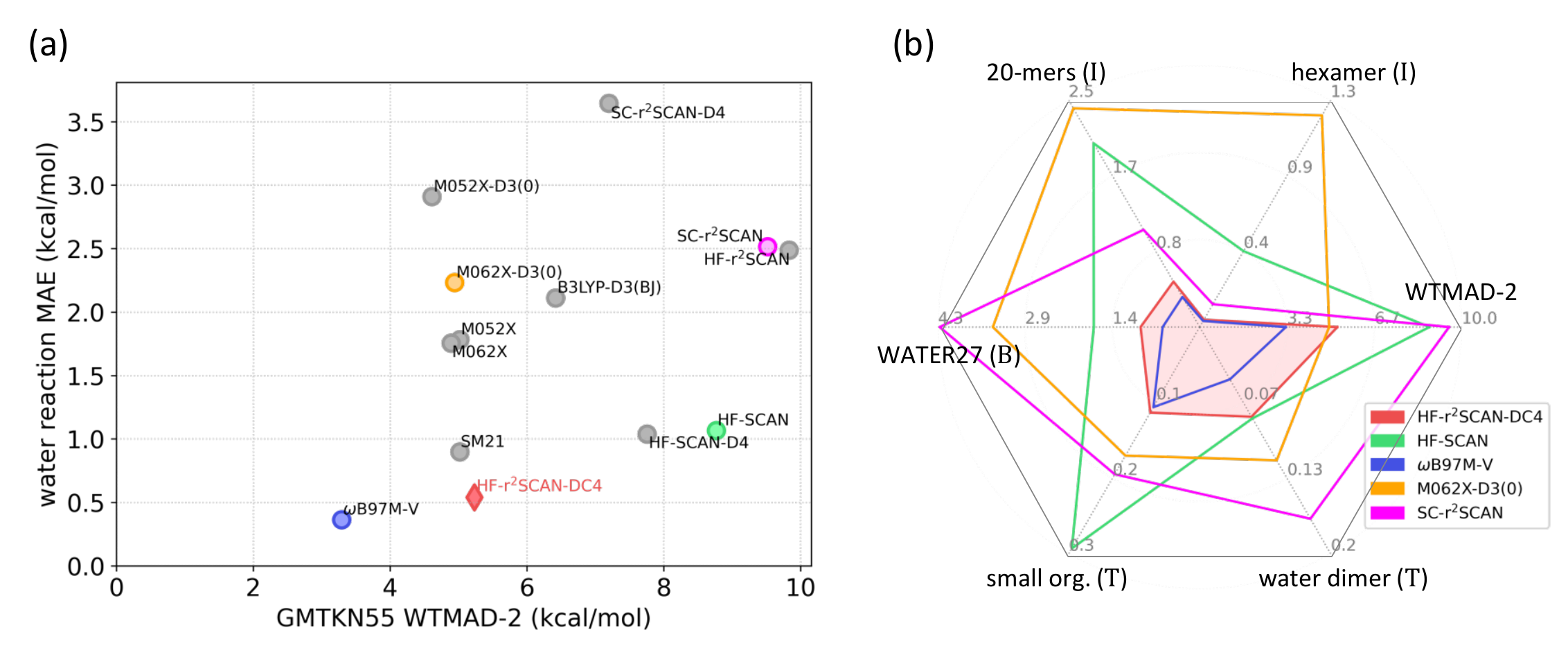}
\caption{
(a) The mean absolute error (MAE) for the 
water-based reactions appear in this work (hexamer isomer energies, water 20-mers isomer energies, WATER27 binding energies, water-small organic molecule interaction energies, and water dimer interaction energies) versus
the weighted-mean-absolute-deviation-2
(WTMAD-2) for the GMTKN55 database for selected functionals.
For a further description of the reactions used in the $y$-axis,  see Section~\ref{sec:water_int} in the supporting information.
HF-SCAN-D4 functional used here is from Ref.~\cite{SM21}.
(b) 
The hexagon plot with MAEs for selected water-based datasets 
and WTMAD-2 values for the whole GMTKN55 databases 
(for WTMAD-2 values for other GMTKN55 database,  see 
Figure~\ref{fgr:g}).
Abbreviations of isomerization (I), binding (B), and interaction (T) energy
are noted in the vertex caption.
MAEs of  HF-r$^2$SCAN-DC4 for individual GMTKN55 datasets are shown in Table~\ref{tbl:gmtkn}.
In Figure~\ref{fgr:nohb},  we give further details about the interaction energies used in the "water-small organic molecule" dataset.
}
\label{fgr:w}
\end{figure*}

IIn Figure~\ref{fgr:wc},
we study the performance of different variations for microhydration of cytosine,  
by specifically focusing on the interaction energies in water$\cdots$cytosine complexes.
We generate these complexes as described in Section~\ref{sec:cyto}, 
and in all of them, water interacts with cytosine 
through the hydrogen bond formed between the hydrogen atom 
in water and the oxygen atom in cytosine.
For each complex,  the errors of HF-r$^2$SCAN-DC4  are small,  
and with the MAE of 0.09 kcal/mol,  it is the best performer in 
Figure~\ref{fgr:wc}.

The errors of HF-SCAN are much smaller here than for cytosine dimers (Figure~\ref{fgr:1}(a)),
in which the role of dispersion is more important. 
Nevertheless,  
HF-r$^2$SCAN-DC4 provides here a significant improvement over HF-SCAN.  
It is also interesting to observe what happens after we add the dispersion correction to 
HF-r$^2$SCAN and its SC counterpart. 
In the case of HF-r$^2$SCAN, the errors in the interaction energies are greatly reduced (roughly by a factor of 6 on average).
In stark contrast,  adding D4 to SC-r$^2$SCAN significantly deteriorates its accuracy,  as
SC-r$^2$SCAN already overbinds water$\cdots$cytosine complexes and D4 makes 
the overbinding stronger.

\subsection{Wide applicability of HF-r$^2$SCAN-DC4}

A functional that works extremely well for pure water but nothing else is not widely applicable.
Recently, GMTKN55 of 55 databases has become a popular benchmark for testing the
accuracy of density functionals for main-group chemistry.  Figure~\ref{fgr:w} has been designed
to illustrate performance of functionals for both pure water and on the GMTKN55 database
simultaneously.   
The water metric ($y$-axis on the left) combines most of the reactions with water used in this
paper, and is carefully defined in the supporting information Section~\ref{sec:water_int}.

Figure~\ref{fgr:w}(a) shows errors on GMTKN55 on the $x$-axis and errors 
on the water metric on the $y$-axis,
each in kcal/mol.  
The $x$-axis ranges from about 3-10 kcal/mol, spanning the performance of modern
approximations for main group chemistry, such as atomization energies.   
The $y$-axis range is much
smaller, running less than 4.0 kcal/mol, reflecting the much smaller magnitude of NCIs in water, and
how high accuracy needs to be in order to have an accurate model for water.   
Here, HF-SCAN sets
a high standard, with a water error near 1.0 kcal/mol (the chemical accuracy claimed in Ref.~\cite{DLPP21}),
while most standard-use functionals cannot compete.   On the other hand, SCAN is designed mainly
to improve materials calculations without the cost of a hybrid functional, and HF-SCAN has a high
error on GMTKN55 (about 9 kcal/mol).  Popular functionals have much smaller GMTKN55 errors,
but perform worse on water.
We also show the many combinations of
HF-r$^2$SCAN-DC4 that do not include all the right ingredients, showing they all perform less well on
water than HF-SCAN.   We finally include 
$\omega$B97M-V functional \cite{MH16}, which might be considered the DFT gold-standard here, with the
smallest errors for both water and main-group chemistry.   
But this range-separated functional with nonlocal correlation functional
is far more expensive to compute than most functionals
,including its own D4 variant,\cite{MH18} and
is less practical for DFT-MD simulations than e.g., SCAN.
We have included it here only to show what is possible in principle with DFT.

\begin{figure*}[htb]
\centering
\includegraphics[width=1.95\columnwidth]{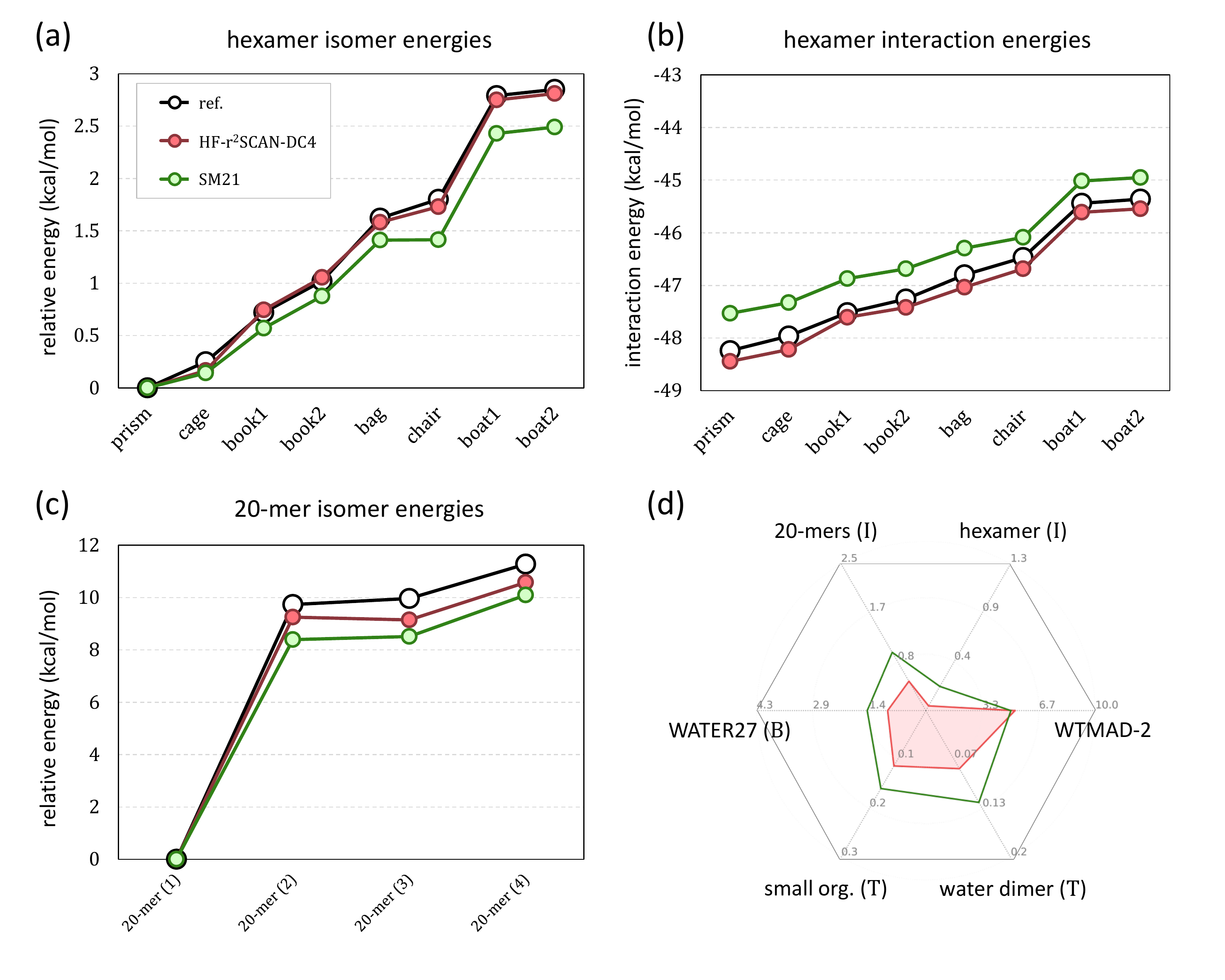}
\caption{
Comparison between HF-r$^2$SCAN-DC4 and
SM21 for (a) water hexamer isomerization energy,
(b) water hexamer interaction energy,
(c) water 20-mer isomerization energy,
and (d) hexgonal plot same as Figure~\ref{fgr:w}(b).
SM21 (green) is HF-r$^2$SCAN but with different D4 parameters obtained in Ref.~\cite{SM21b}.
}
\label{fgr:golo}
\end{figure*}

But the performance of HF-r$^2$SCAN-DC4 is remarkable.  Its errors on both water {\em and} the GMTKN55
dataset are almost half of those of HF-SCAN.   No other functional in our collection comes close for
water.   Clearly, all the chemically-inclined approximations which are comparable for main-group
chemistry do much worse.

In Figure~\ref{fgr:w}(b),  we show the hexagon plots
comparing the MAEs of several density functionals, 
where the position of five vertices denote the MAEs for individual water-based datasets,
while the sixth vertex denotes the overall performance of the functionals for the  whole GMTKN55 databases,  
as measured by 
the weighted-mean-absolute-deviation-2
(WTMAD-2).
It is the MAE for all the reactions from these five water-based datasets that
we use as the quantity on the $y$-axis in Figure~\ref{fgr:w}(a).
The size of the hexagon of HF-r$^2$SCAN-DC4
is the closest to that of more costly  $\omega$B97M-V.
We can also see that the performance of HF-r$^2$SCAN-DC4 is far superior to that of HF-SCAN.
M062X-D3(0),  a meta-hybrid that is very accurate for small organic molecules,\cite{GHBE17}  
and yields WTMAD-2 which is slightly lower than that of 
HF-r$^2$SCAN-DC4.
But,  for water simulations,  
M062X-D3(0) is nowhere close to HF-r$^2$SCAN-DC4,  as can be seen from the position of the remaining five vertices.

\section{Methods}
\label{sec:met}

The basic principles of DC-DFT are covered elsewhere 
in the literature\cite{KSB13,VSKS19}, 
and reviewed in the supporting information.
In most KS-DFT calculations, the error in the density has a negligible effect on the energy errors.
But sometimes the error in a SC density leads to a noticeable contribution, which can
be reduced if a more accurate density is used instead.  For many semilocal
exchange-correlation approximations in molecular
calculations, when a calculation is density sensitive, often the HF density then yields significantly
smaller energy errors.   These principles have led to improved energetics in reaction barrier heights, 
electron affinities, and also for the ground state geometries of noncovalent interaction systems, etc.\cite{KSB14,JS08,LFB10,LB10}

Application of the principles of DC-DFT is subtle in the case of r$^2$SCAN-D4, because of the need to separate
out the error due to density correction from the fitting of the D4 corrections.   
For example, for halogen bonds,
the density-driven errors are far larger than dispersion corrections, so all fitting must be done on density-corrected
energetics.  Moreover, when empirical functionals contain parameters, such parameters should be fit only on
density-insensitive calculations, so that the parameters optimize the `true' functional error.

With these principles in mind, we find the parameters for HF-r$^2$SCAN-DC4 
using the density-insensitive calculations in the GMTKN55 dataset
as a training set 
while using water$\cdots$water pair interaction energy as a validation set.
We find their optimum values by minimizing MAE values over all such
cases.   
This is detailed in the supporting information section~\ref{sec:d4para}.   
This is why we use the acronym DC4 instead of D4, meaning 
that we use the principles of DC-DFT to find the underlying D4 parameters.

In Refs.~\cite{SVSB22} and \cite{SSVB22},
we proposed DC(HF)-DFT,  
a DC-DFT procedure 
that discriminately uses HF 
densities based on the  density sensitivity criterion.
The main idea of DC(HF)-DFT is to use HF-DFT for density-sensitive (DS) reactions and SC-DFT for density-insensitive (DI) reactions
 (possible spin-contaminations of the HF results are also taken into account as detailed in Section~\ref{sec:hfdft}). 
While we consider DC(HF)-DFT a state-of-the-art DC-DFT procedure, 
for our HF-r$^2$SCAN-DC4  we use HF-DFT,
meaning that the functional is always evaluated on the HF density regardless of the 
sensitivity criterion.
To use DC(HF)-DFT,  we need to compute density sensitivity for each reaction of interest and possibly make adjustments to its cutoff value which is used to declare whether a given reaction is DS or DI.\cite{SVSB22,SSVB22}  This would also require having two sets of D4 parameters,  one for DS and 
the other
for DI reactions.
All these efforts 
would undermine
the ease of use of r$^2$SCAN,  which is a general-purpose functional.
For this reason and encouraged by the very good performance of HF-DFT with SCAN-like functionals\cite{SRP15,FKNP20}, 
we employ HF-DFT\cite{TCCL} as a DC-DFT procedure for HF-r$^2$SCAN-DC4.
While our HF-r$^2$SCAN-DC4 can be routinely used by applying it to HF orbitals without ever needing to calculate density sensitivity of a given reaction,  
the use of DC-DFT principles and density sensitivity 
is vital for our training of HF-r$^2$SCAN-DC4 as explained above. 

To illustrate what can happen when these principles are not applied, we show results from Ref.~\cite{SM21b}.
This is a version of HF-r$^2$SCAN-D4, but where all reactions in GMTKN55 were used, and the
WTMAD-2\cite{GHBE17} was used as
the cost function instead.
Figure~\ref{fgr:golo} illustrates the results for the larger water clusters.  In every case, they are noticeably
worse than ours.   Moreover, (d) shows that, apart from matching on WTMAD-2 measure,
HF-r$^2$SCAN-DC4 yields more accurate results in every other case.

\section{Conclusions}
\label{sec:conc}

The work of Ref.~\cite{DLPP21} was a breakthrough in models for water, 
showing that, by using the principles of DC-DFT,
a moderate-cost density functional approximation approached chemical accuracy 
for many relevant properties
of small water clusters.   
However that functional is lacking in dispersion corrections, yielding large errors for energetics
between organic and biological molecules.   
It also inherits some of the numerical issues of the original SCAN
functional, which have been eliminated by using r$^2$SCAN instead in most other applications.  However,
the small differences between these two wreak havoc on the much smaller scale of subtle energy differences
of water clusters.

The present work shows that, by a very careful application of the principles of DC-DFT, all these difficulties
can be overcome, and even greater accuracy achieved for pure water, while still including dispersion for other
molecules where it can be vital.   Finding the correct parameters depends crucially on training on only
density-insensitive chemical reactions, as inclusion of
density-sensitive reactions yields suboptimal values for the parameters.

Even if HF-r$^2$SCAN-DC4 could be run at close to meta-GGA cost, KS-DFT MD simulations are typically far more costly than MD with machine learning (ML) interatomic potentials. 
But accurate force-field generation requires highly accurate reference energetics data as a training set, and CCSD(T) or 
Quantum Monte Carlo
(QMC) methods are frequently used as reference methods these days.\cite{MBP14}
Due to the large computational cost for such {\it ab initio} calculation, a more practical yet accurate method is in demand, and HF-r$^2$SCAN-DC4 can replace them for calculating moderately large biomolecular systems.
We suggest HF-r$^2$SCAN-DC4 be tested and applied in solution wherever practical.
\\

\acknowledgement{
ES, SS, YK, and HY are grateful for support from the National Research Foundation of Korea (NRF-2020R1A2C2007468 and NRF-2020R1A4A1017737)
and
Samsung Electronics (IO211126-09176-01). 
KB acknowledges funding from NSF (CHE-2154371).
SV acknowledges funding from the Marie Sk\l{}odowska-Curie grant 101033630 (EU’s Horizon 2020 programme).
We thank John Perdew and Francesco Paesani and his group for many useful discussions.
}


\newcommand{\beginsupplement}{%
        \setcounter{table}{0}
        \renewcommand{\thetable}{S\arabic{table}}%
        \setcounter{figure}{0}
        \renewcommand{\thefigure}{S\arabic{figure}}
         \renewcommand{\thesection}{S\arabic{section}}%
          \setcounter{section}{0}
     }

\beginsupplement

\clearpage

\textbf{Supporting Information}

\section{DFT and DC-DFT}
\label{sec:dft}

In KS-DFT \cite{KS65},  the ground state energy 
is obtained by minimizing the following functional over densities,
\begin{equation}
\label{eq:gs}
E_v [n]= T_{\s}[n] + U_{\sss H}[n]  + E\xc[n] +  \intr n(\br)v(\br),
\end{equation}
where the minimizing $n_v(\br)$ is the ground-state density,  $T_{\s}[n]$ is the KS noninteracting kinetic energy functional,  $U_{\sss H}[n]$ the Hartree energy, and $E_{\xc}[n]$ is the exchange-correlation (XC) functional. 
In practical KS calculations,  the only term that of Eq.~\ref{eq:gs} that is approximated is $E_{\xc}[n]$.
Whatever approximation we use for XC,  it yields an error in both energies and the densities,
as the minimizing density in Eq.~\ref{eq:gs} is different from the exact when an approximate XC functional is used in place of its exact counterpart. 
We can write the error of any 
KS calculation as: $\Delta E = \tilde E[\tilde \n]-E[n]$ as,  where $E [n]$ is the exact functional (Eq. ~\ref{eq:gs}) and $n$ is the exact density (we drop the $v$ subscript for brevity),  while tildes denote approximate quantities. 
The main idea of DC-DFT is to separate the errors in $\Delta E$ into a functional error,  which is present even with the exact density \cite{KSB13}:
\ben
\label{eq:f}
\Delta E\f = \tilde E [\n] - E[\n] = \tilde E\xc[\n]-E\xc[\n],
\een
and the density-driven error is the remainder of $\Delta E$:
\ben
\label{eq:d}
\Delta E\d= \tilde E[\tilde\n]-\tilde E[\n].
\een

In most KS-DFT calculations,  $\Delta E\f$ strongly dominates $\Delta E$,
implying that approximate KS densities are very good as measured by the impact on energies.
A very simply example for such case is the total energy of the helium atom,  whose energy computed from the PBE functional\cite{PBE96} barely changes after the PBE density is replaced with the exact one.\cite{KSB13}
However, for a significant number of chemical domains (e.g.,  anions and barrier heights),  $\Delta E\d$
can be much larger than $\Delta E\f$\cite{VSKS19,NSSB20}.   
A good example for such a case is H$^-$ with the PBE functional,  which 
gives excellent energies when evaluated on the exact densities,  whereas the self-consistent PBE density cannot even bind two electrons for this anion.

\section{Spotting and curing large density-driven errors}
\label{sec:dde}

The following two questions are of key importance in DC-DFT:
(i) How do we spot cases where $\Delta E\d$ is large?; 
(ii) What do we do in such cases to reduce large density-driven errors? 
With access to exact densities we can 
easily answer (i),  as with those we can easily measure $\Delta E\d$ by Eq.  \ref{eq:d}. 
At the same time,  we answered (ii),  as $\Delta E\d$  entirely vanishes with exact densities.
However,  exact densities are available only for small systems \cite{KSB13,SSB18},  and 
if we always had 
access to exact energies and densities from 
highly accurate wavefunction theories,  we would not even bother with KS-DFT. 
Thus one needs to find more practical ways to answer (i) and (ii).  
In relation to (i),  the following quantity has been introduced:
\ben
\tilde S = \left| \tilde E[\n\LDA] - \tilde E[\n\HF] \right|,
\label{eq:s}
\een
and is called {\em density sensitivity}.  
$\tilde S$ requires two nonempirical densities: the HF densities which are typically overlocalized and 
the local density approximation (LDA) densities which are typically delocalized.  
$\tilde S$ serves as a practical measure of density sensitivity 
of a given reaction and
approximate functional. 
For small molecules,  $\tilde S$ greater than the heuristic cutoff of 2 kcal/mol implies density sensitivity,  indicating that the calculation may suffer from a large  $\Delta E\d$.
What should we do then to reduce large $\Delta E\d$?
In these cases,  
evaluating an approximate functional on the HF in place of self-consistent densities will likely reduce $\Delta E\d$ and likely improve the functional's performance.  
This procedure,  called HF-DFT,  is the practice of evaluating an XC approximation on the HF density and orbitals. It had been used a long before DC-DFT 
was proposed \cite{CC90,GJPF92a,GJPF92b,S92,OB94}
but only because HF densities were more convenient than self-consistent densities.
KS-DFT does not always benefit from HF densities (e.g,  cases where $\Delta E\d$ is small) and in Refs.  \cite{SVSB21,SVSB22,SSVB22} 
we discuss in more details formal and practical 
(dis)advantages of 
HF-DFT over SC-DFT.  

\section{HF-DFT and related DC-DFT procedures}
\label{sec:hfdft}

Following the idea that in some cases HF-DFT works better and SC-DFT in others, 
DC(HF)-DFT has been proposed \cite{SVSB22,SSVB22}.
It is a procedure that discriminately uses HF densities,  as DC(HF)-DFT 
becomes HF-DFT for
cases that are both density-sensitive 
($\tilde{S}$ above a given cut-off value, 2 kcal/mol as discussed in Ref.~\cite{SSB18}.) 
and whose HF solution is not severely spin-contaminated.  Otherwise,  
DC(HF)-DFT  reverts to SC-DFT.
Since HF-DFT uses the HF density as a proxy for the exact density, 
we only use it when there is little or no spin contamination. 
We calculate the expectation values of the spin-squared operator, $S^2$, 
and only use the HF density if the <$S^2$> from the HF calculation deviates less than 10\% 
from the exact <$S^2$> as discussed in Refs.~\cite{SVSB22} and \cite{RHBH20}.  
Otherwise, we use the self-consistent density.
As it combines the best of both of them,  DC(HF)-DFT comes with a range of advantages over both HF-DFT and SC-DFT as further detailed in Ref.~\cite{SVSB22}. 
These advantages come at a small extra cost, as for DC(HF)-DFT we need to run up to three distinct self-consistent cycles to obtain the three densities (the SC density for a given functional and those from HF and LDA needed to calculate $\tilde{S}$).  
While we consider DC(HF)-DFT the state-of-the-art DC-DFT-based procedure,
in the present work we want a simple DC framework that can be applied easily and routinely.  
r$^2$SCAN and SCAN are general-purpose functionals
and the ease of their use would be undermined if tandem with DC(HF)-DFT,  which would require always calculating $\tilde{S}$ and possibly making adjustments to its cut-off value.
For this reason and encouraged by the very good performance of HF-DFT with SCAN-like functionals\cite{SRP15, FKNP20}, 
we employ HF-DFT as a DC-DFT procedure throughout this work. 
 As said,  constructing the robust and accurate  HF-r$^2$SCAN-DC4
 is the central objective of this work.
While the resulting HF-r$^2$SCAN-DC4 can be routinely used by applying it to HF orbitals without ever needing to calculate $\tilde{S}$ of a given reaction,  
the use of $\tilde{S}$ 
is vital for our training of HF-r$^2$SCAN-DC4.
Specifically,   
we use density-sensitivities of the training reactions to fit the D4 part of HF-r$^2$SCAN.
Further technical details of this fitting procedure will be given in the next section.

\section{Optimizing dispersion parameters}
\label{sec:d4para}
D4 stands for the 
generally applicable atomic-charge dependent London dispersion correction term
developed by Grimme and co-workers.\cite{CEHN19}.
It has 4 functional-dependent parameters
$s_6$, $s_8$, $a_1$, and $a_2$.
Following Refs.~\cite{GAEK10} and~\cite{CEHN19},
we set $s_6$ to unity as is common for functionals that do not
capture long-range dispersion interactions. 
We optimized the $s_8$, $a_1$, and $a_2$ parameters 
by minimizing the mean absolute error (MAE) 
for the density-insensitive GMTKN55 reactions by following the 
DC-DFT ideas of Ref.~\cite{SVSB21}.
However, the density-insensitive reactions in GMTKN55 largely fall into two distinct parameter groups 
for HF- r$^2$SCAN: $s_8$ has a negative value for noncovalent interactions, 
but is positive for the rest.   
The difference in MAE of density-insensitive cases between those two groups is miniscule (below 0.01 kcal/mol). 
For example, ($s_8$,$a_1$,$a_2$)=(-0.20,0.07,6.50) gives 1.209 kcal/mol for the density-insensitive MAE 
while (0.39,0.09,7.02) gives 1.210 kcal/mol.   
Such a difference is not meaningful.  
Small changes in computational details such as DFT grid information, two-electron operator fitting scheme, etc. 
changes the values of the parameters, 
since reaction energy errors and density sensitivity values can be changed by 0.01 kcal/mol with those changes.   
To eliminate this ambiguity while ensuring accuracy in water interactions, 
we include the density-insensitive water$\cdots$water pair interaction energy 
as a validation set. 
The two most stable water hexamers, the prism and the cage, are used to calculate 
the water$\cdots$water 2-body interaction energy error per dimer, 
relative to CCSD(T)/CBS in Ref.~\cite{RSBP16}.   
We multiply its weight by 7 in our loss function to produce a better defined minimum 
and regularize the result (if we used 1, it has no effect; if we used 1000, we simply fit to this data).   
We can rationalize this value by noting that the mean density sensitivity of these pairs is 0.27 kcal/mol, 
which is about 1/7th of our density sensitivity cutoff.
The resulting values for the three parameters are: 
-0.36,  0.23, 5.23
for $s_8$, $a_1$, and $a_2$ each.

\section{Additional results for the GMTKN55 database}

In Table~\ref{tbl:gmtkn},  we list MAE (kcal/mol)
of SC-r$^2$SCAN and HF-r$^2$SCAN functional
with and without the dispersion correction for the chemically diverse GMTKN55 database \cite{GHBE17}.
def2-QZVPPD basis set is used.

\begin{table*}[htb]
\centering
\resizebox{0.5\textwidth}{!}{
\begin{tabular}{lcccc}
          & HF-r$^2$SCAN & HF-r$^2$SCAN-DC4 & SC-r$^2$SCAN & SC-r$^2$SCAN-D4 \\ \hline \hline
MAE (all)  & 2.82     & 2.42        & 2.98    & 2.91        \\
MoM (all)  & 2.77     &2.32        & 3.24    & 3.10       \\
WTMAD-2 (all)$^\dagger$  & 8.54     &5.18        & 8.65    & 7.10        \\ \hline
basic$^\dagger$  &4.19 &4.09 &4.92 &4.89  \\
react.$^\dagger$  &8.10 &5.85 &9.48 &8.11  \\
barriers$^\dagger$  &7.18 &8.04 &13.11 &13.56  \\
inter. NCI$^\dagger$  &13.27 &4.88 &10.96 &6.77  \\
intra. NCI$^\dagger$  &11.96 &4.80 &8.67 &5.87  \\ \hline
\multicolumn{5}{c}{basic}                                          \\
W4 &6.92 &6.41 &3.81 &3.86  \\
G21EA &4.59 &4.55 &3.88 &3.86  \\
G21IP &4.56 &4.54 &4.66 &4.64  \\
DIPCS10 &4.42 &4.34 &5.18 &5.14  \\
PA26 &1.85 &1.80 &2.44 &2.40  \\
SIE4x4 &12.14 &12.24 &17.93 &17.98  \\
ALKBDE10 &4.71 &4.66 &5.01 &5.00  \\
YBDE18 &3.89 &3.67 &3.89 &3.36  \\
AL2X6 &1.02 &1.23 &0.93 &1.58  \\
HEAVYSB11 &5.60 &4.32 &3.91 &3.15  \\
NBPRC &1.24 &1.24 &1.60 &1.52  \\
ALK8 &1.73 &1.77 &2.73 &2.88  \\
RC21 &2.82 &2.39 &4.57 &4.95  \\
G2RC &4.24 &4.56 &5.38 &5.55  \\
BH76RC &2.57 &2.56 &2.97 &2.97  \\
FH51 &1.71 &1.72 &2.19 &2.16  \\
TAUT15 &1.11 &1.10 &1.58 &1.57  \\
DC13 &8.96 &7.97 &8.63 &7.72  \\
\multicolumn{5}{c}{react.}                                          \\
MB16-43 &10.48 &10.79 &12.59 &14.08  \\
DARC &3.82 &2.00 &3.71 &2.70  \\
RSE43 &0.96 &0.96 &1.55 &1.51  \\
BSR36 &3.20 &0.16 &2.32 &0.48  \\
CDIE20 &1.21 &1.13 &1.63 &1.61  \\
ISO34 &1.52 &1.36 &1.36 &1.29  \\
ISOL24 &4.34 &3.01 &4.96 &4.10  \\
C60ISO &3.52 &3.88 &5.35 &5.57  \\
PArel &1.14 &1.15 &1.55 &1.54  \\
\multicolumn{5}{c}{barriers}                                          \\
BH76 &2.85 &2.90 &6.87 &6.98  \\
BHPERI &4.14 &5.95 &3.86 &4.65  \\
BHDIV10 &3.49 &3.89 &5.98 &6.11  \\
INV24 &1.35 &1.34 &1.22 &1.14  \\
BHROT27 &0.62 &0.64 &0.76 &0.76  \\
PX13 &4.74 &5.03 &8.75 &8.83  \\
WCPT18 &2.73 &3.18 &5.81 &5.99  \\
\multicolumn{5}{c}{inter. NCI}                                          \\
RG18 &0.26 &0.10 &0.23 &0.16  \\
ADIM6 &2.65 &0.45 &1.98 &0.34  \\
s22 &1.55 &0.42 &1.18 &0.24  \\
S66 &1.42 &0.30 &1.02 &0.26  \\
HEAVY28 &0.71 &0.37 &0.52 &0.30  \\
WATER27 &3.94 &1.01 &4.24 &6.30  \\
CARBHB12 &0.63 &0.60 &0.88 &1.06  \\
PNICO23 &0.71 &0.29 &0.64 &0.76  \\
HAL59 &1.01 &0.41 &0.99 &0.80  \\
AHB21 &0.63 &0.68 &1.15 &1.35  \\
CHB6 &0.46 &0.58 &0.49 &0.52  \\
IL16 &1.88 &0.43 &0.33 &0.64  \\
\multicolumn{5}{c}{intra. NCI}                                          \\
IDISP &6.83 &1.58 &10.67 &7.21  \\
ICONF &0.30 &0.23 &0.32 &0.29  \\
ACONF &0.54 &0.16 &0.38 &0.18  \\
Amino20x4 &0.40 &0.27 &0.26 &0.19  \\
PCONF &1.22 &0.44 &1.05 &0.41  \\
MCONF &0.94 &0.21 &0.63 &0.45  \\
SCONF &0.57 &0.19 &0.37 &0.51  \\
UPU23 &1.18 &0.38 &0.95 &0.41  \\
BUT14DIOL &0.40 &0.14 &0.14 &0.23  \\ \hline \hline
\end{tabular}
}
\caption{
Mean-absolute-error (MAE)
of GMTKN55 for selected functionals for the individual datasets in the GMTKN55 database.
WTMAD-2 value from Ref.~\cite{GHBE17} and mean of means (MoM) values are 
also noted. WTMAD-2 values for individual GMTKN55 categories 
$^\dagger$(basic properties and reaction energies for small systems (basic), reaction energies for large systems and isomerisation reactions (react.), reaction barrier heights (barriers), intermolecular noncovalent interactions (inter. NCI), and intramolecular noncovalent interactions (intra. NCI)) are also added.
All units are kcal/mol.
}
\label{tbl:gmtkn}
\end{table*}

\section{Additional results for water clusters}
\label{sec:water}

\subsection{MD generated dimer structures}
The structures used in Figures~\ref{fgr:1}(e) and \ref{fgr:dimer}(b),
have been obtained from the 
molecular dynamics (MD) simulation.
The simulation is performed within the XTB package\cite{BEG19}
and the GFN-FF force field\cite{SG20},  enabling us to generate the various dimer configurations.
The total simulation time is 50 ps,
while the integration time step is 4.0 fs
using a Berendsen thermostat at 298K
in the NVT ensemble.
We use the SHAKE algorithm to constrain bonds, 
for all bonds with 4 amu for the hydrogen atom mass.
Then,  we randomly selected 110 different
configurations for the water$\cdots$water dimer
and 80 for the water$\cdots$Aspirin dimers.
The reference interaction energies are then calculated with
DLPNO-CCSD(T)-F12/TightPNO method 
with the aug-cc-pvqz basis set for water$\cdots$water dimers
and aug-cc-pvtz basis set for water$\cdots$Aspirin dimers.

\subsection{Many-body expansion of the interaction energy}

The interaction energy can be decomposed into 
2-body, 3-body, etc. by using the many-body expansion.\cite{GPCS11}
For example,  the interaction energy of the water hexamer 
can be divided into $K$-body contributions,
\begin{equation}
E_{int} = E_{int}^{2-body} + E_{int}^{3-body} + \cdots + E_{int}^{6-body}
\end{equation}
where $E_{int}^{K-body}$ is the K-body interaction energy
which can be calculated from the total energy of the 
subcluster of the $N$-mer cluster: \cite{GPCS11}
\begin{equation}
E_{int}^{K-body}=\sum_{i=1}^K(-1)^{K-i}
\begin{pmatrix}
  N-i \\
  K-i
\end{pmatrix}
S_{tot}^i
\end{equation}
where $S_{tot}^i$ stands for the total energy summation of the $i$-th
monomer subcluster.

\section{Additional results for the complexes with cytosine}
\label{sec:cyto}
In Figure~\ref{fgr:wc},
we plot the cytosine$\cdots$water and cytosine$\cdots$cytosine
interaction energy error plots,  respectively.
The 14 different cytosine$\cdots$cytosine configurations and
reference interaction energies are from Ref.~\cite{KS19}.
For the cytosine$\cdots$water interaction,
we place two water molecules around 14 different cytosine$\cdots$cytosine
structures
and optimized the water molecular coordinates while
fixing the cytosine$\cdots$cytosine coordinates.
B3LYP functional is used for the geometry optimization.
DLPNO-CCSD(T)-F12/aug-cc-pvqz with TightPNO is used as a reference 
cytosine$\cdots$water interaction energy with the ORCA package.\cite{ORCA}

\section{Interactions including water}
\label{sec:water_int}

\begin{figure*}[htb]
\centering
\includegraphics[width=1.95\columnwidth]{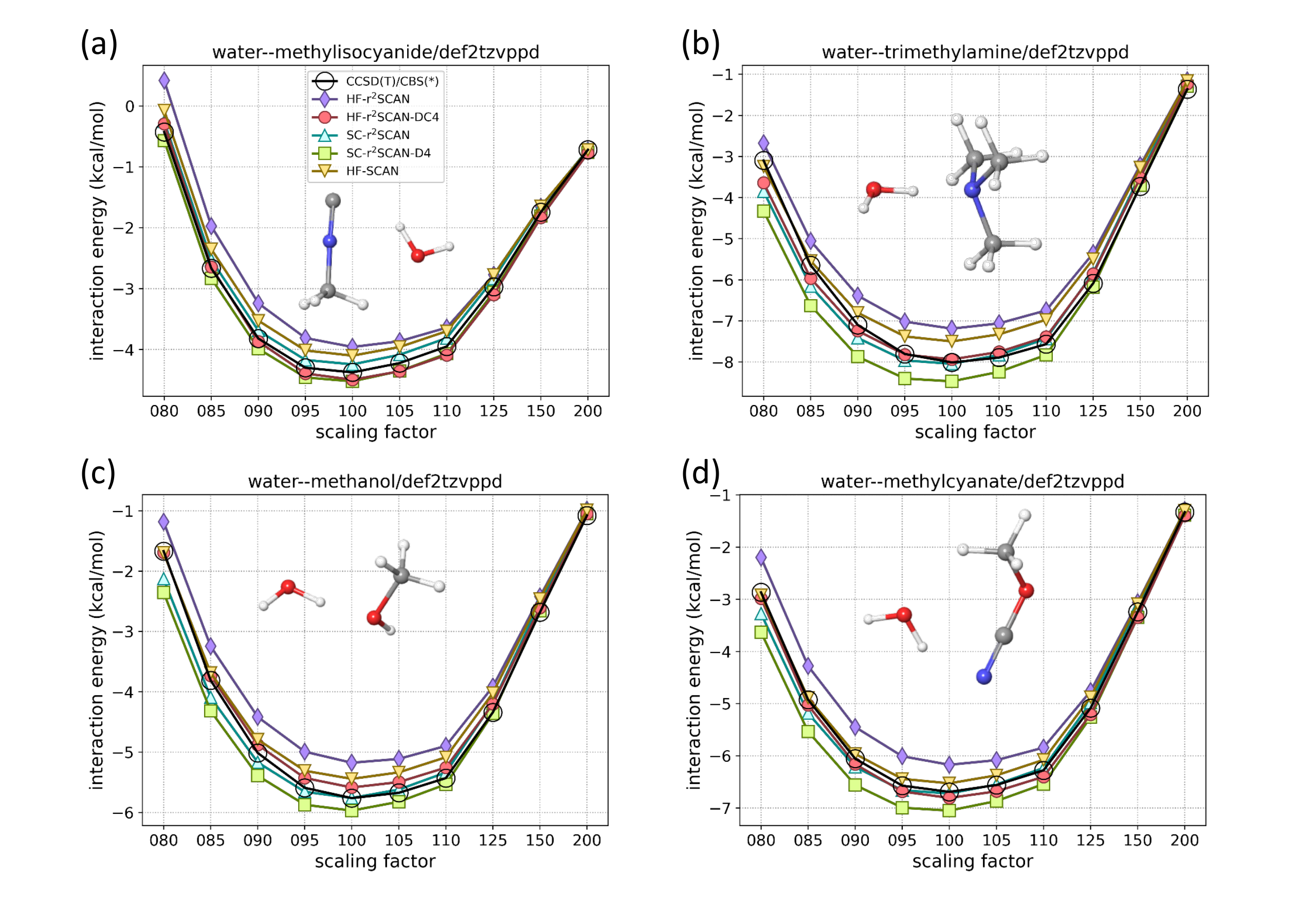}
\caption{
Interaction energy curve
for two non-hydrogen bonded systems.
Geometries are from the HB375 dataset described in Ref.~\cite{R20}.
The x-axis show the scaled distance factor, $f$,
which is used to make a translated vector $t$,
$t=\frac{v}{|v|}(f-1)r_{ref}$
where $v$ is the bond direction vector
and $r_{ref}$ is the distance between
the hydrogen and the electron-donor atom.
Detailed information about the x-axis can be found in
Ref.~\cite{R20}.
}
\label{fgr:nohb}
\end{figure*}

For the calculations shown in Figure~\ref{fgr:w}(a),
we combined energies of 145 reactions involving water. 
They include:
the water hexamer isomerization in Figure~\ref{fgr:1}(b),
the water binding energy of WATER27 dataset in Figure~\ref{fgr:1}(c),
the water 20-mer isomerization in Figure~\ref{fgr:1}(d),
the water dimer stationary point geometry interaction energy in Figure~\ref{fgr:dimer}(a),
and water$\cdots$small organic molecule interaction energy in Figure~\ref{fgr:nohb}.


\clearpage

\begin{figure*}[h]
\centering
\includegraphics[width=0.95\columnwidth]{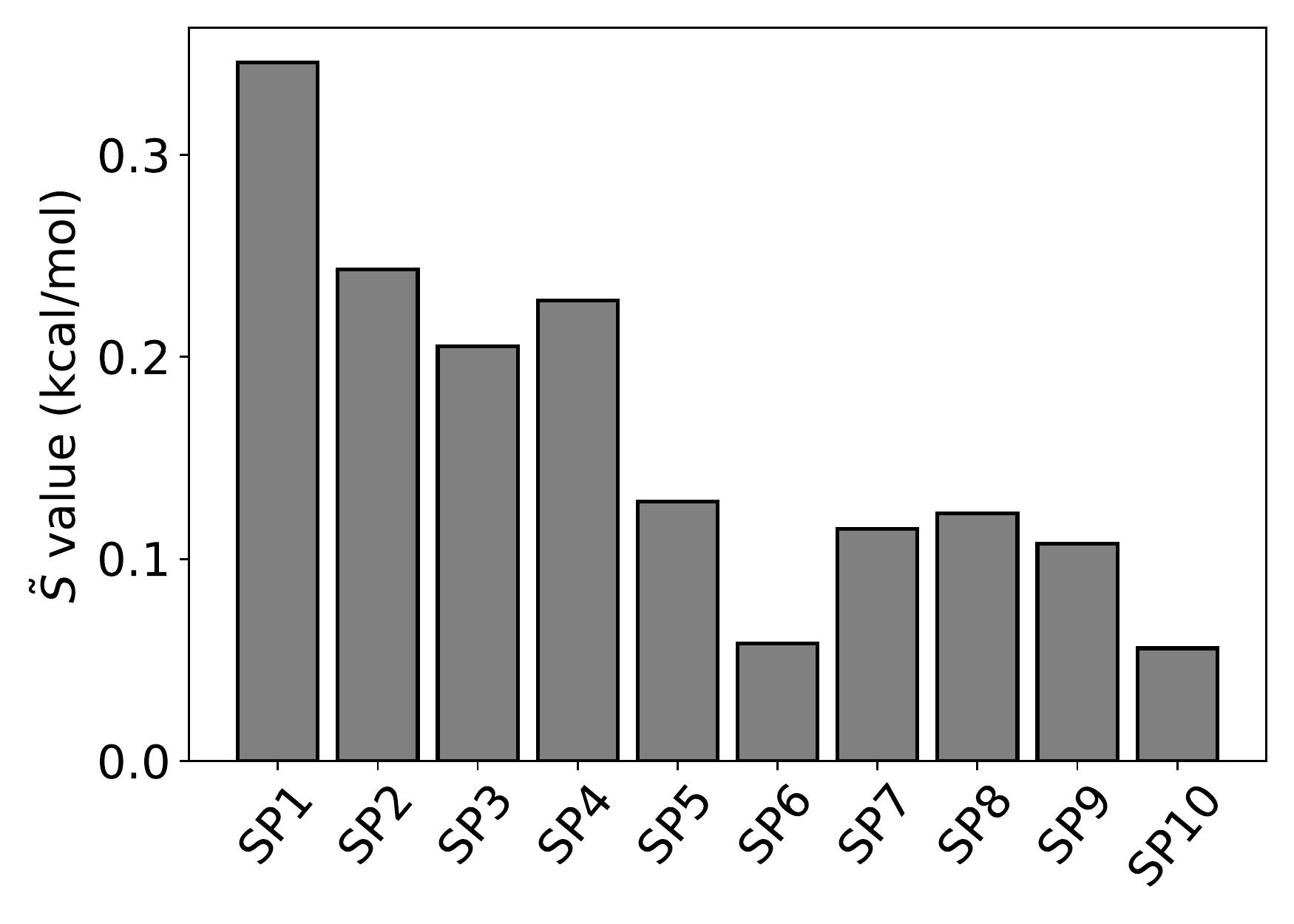}
\includegraphics[width=0.95\columnwidth]{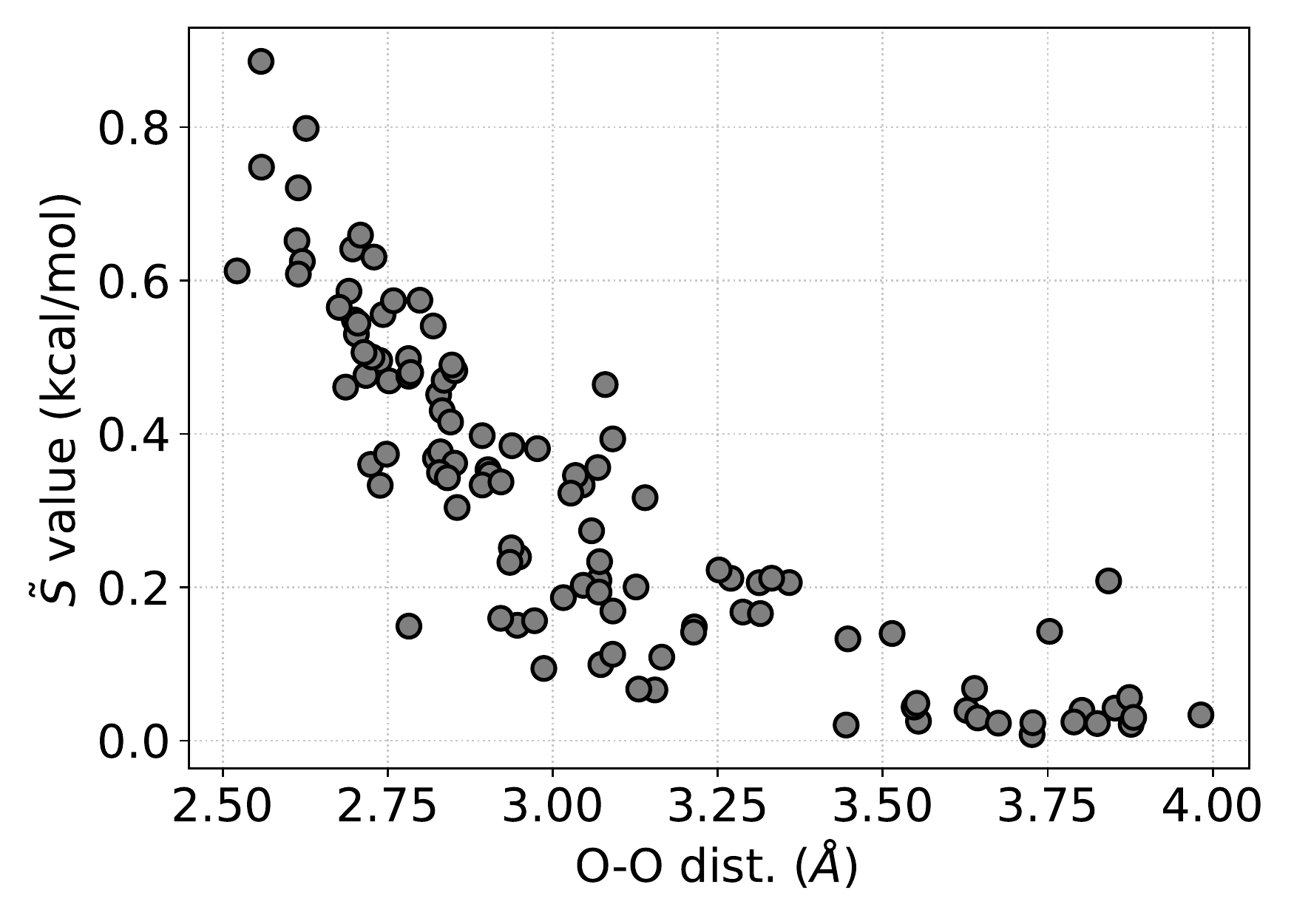}
\caption{
Density sensitivity plot of 
(left) Smith dimer configuration and (right) MD generated dimer configuration
in Figure~\ref{fgr:dimer}.
}
\label{fgr:dimer_s}
\end{figure*}

\begin{figure*}[h]
\centering
\includegraphics[width=0.95\columnwidth]{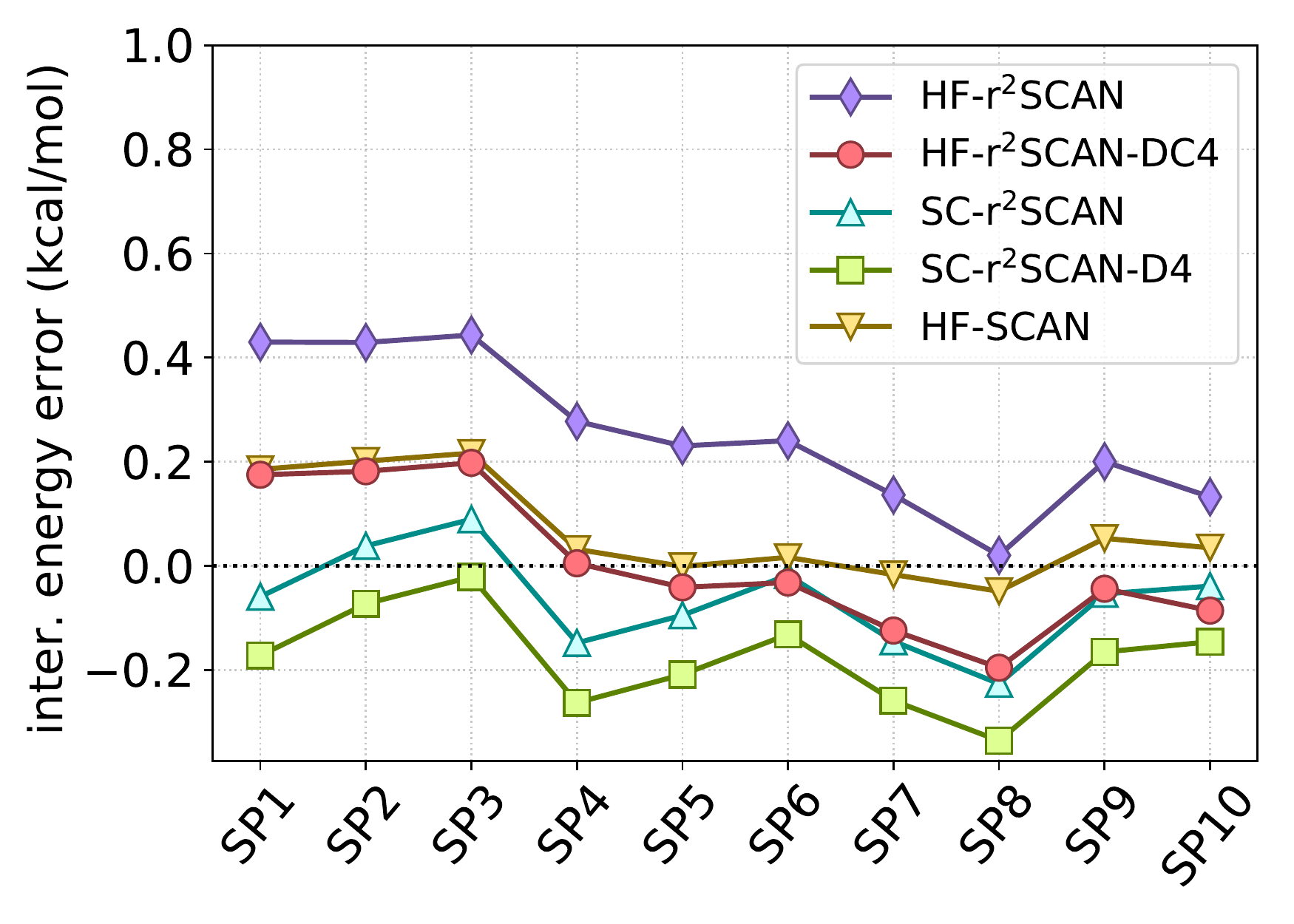}
\includegraphics[width=0.95\columnwidth]{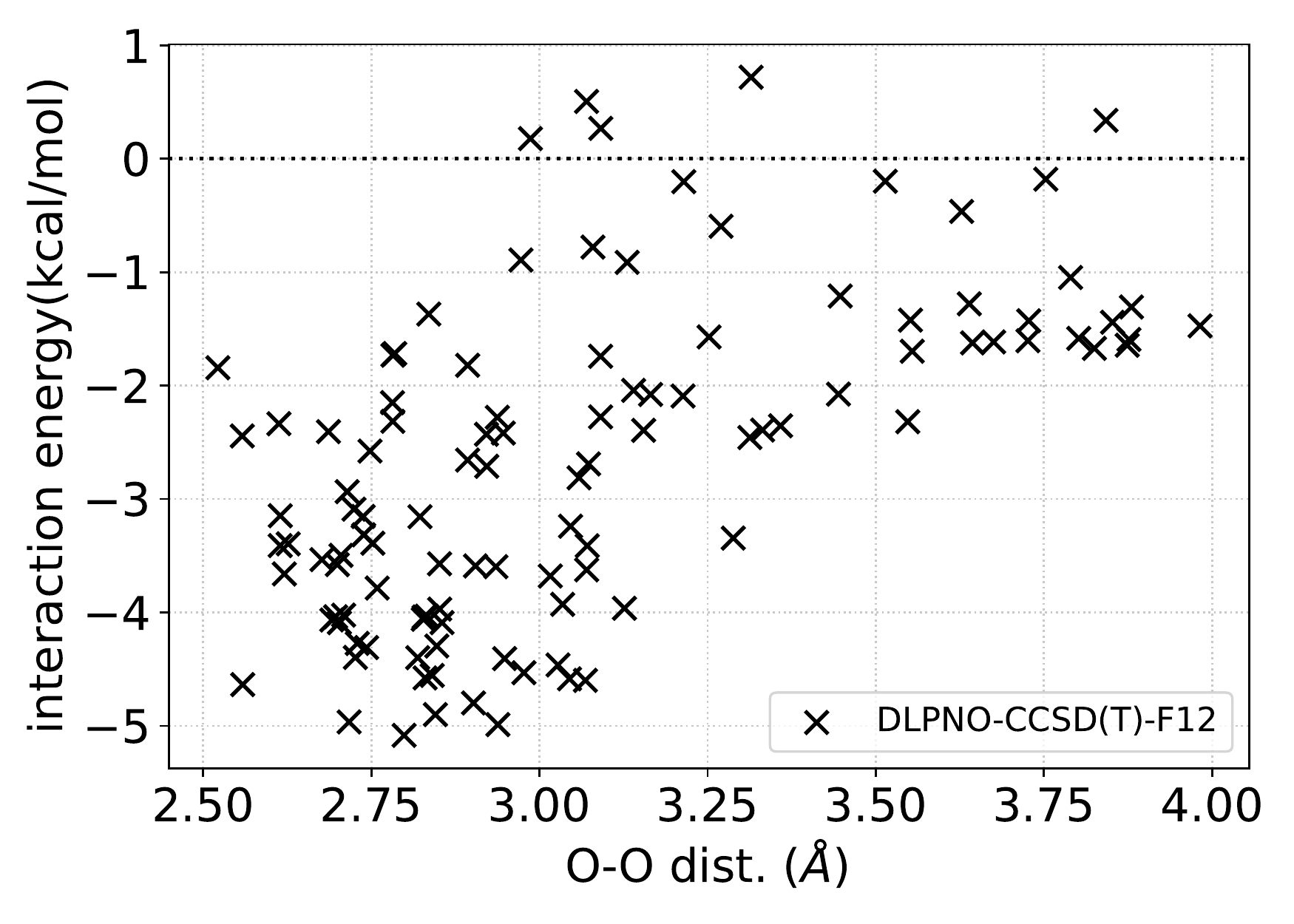}
\caption{
(left) Smith dimer interaction energy errors and
(right) reference dimer interaction energies of MD generated dimer
corresponding to Figure~\ref{fgr:dimer}.
}
\label{fgr:dimer_err}
\end{figure*}

\begin{figure}[htb]
\centering
\includegraphics[width=0.95\columnwidth]{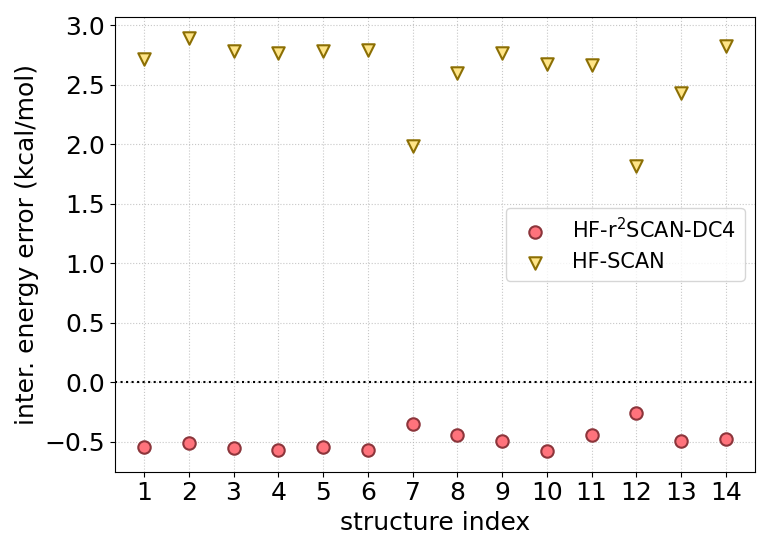}
\caption{
Interaction energy error plot of cytosine$\cdots$cytosine compounds
for HF-r$^2$SCAN-DC4/aug-cc-pvqz and HF-SCAN.
Reference
HF/7Z-CP+MP2/CBS(6,7)-CP+dCC(cc-pVTZ-F12)
energies are from Ref.~\cite{KS19}.
}
\label{fgr:ks19_err}
\end{figure}

\begin{figure}[htb]
\centering
\includegraphics[width=0.95\columnwidth]{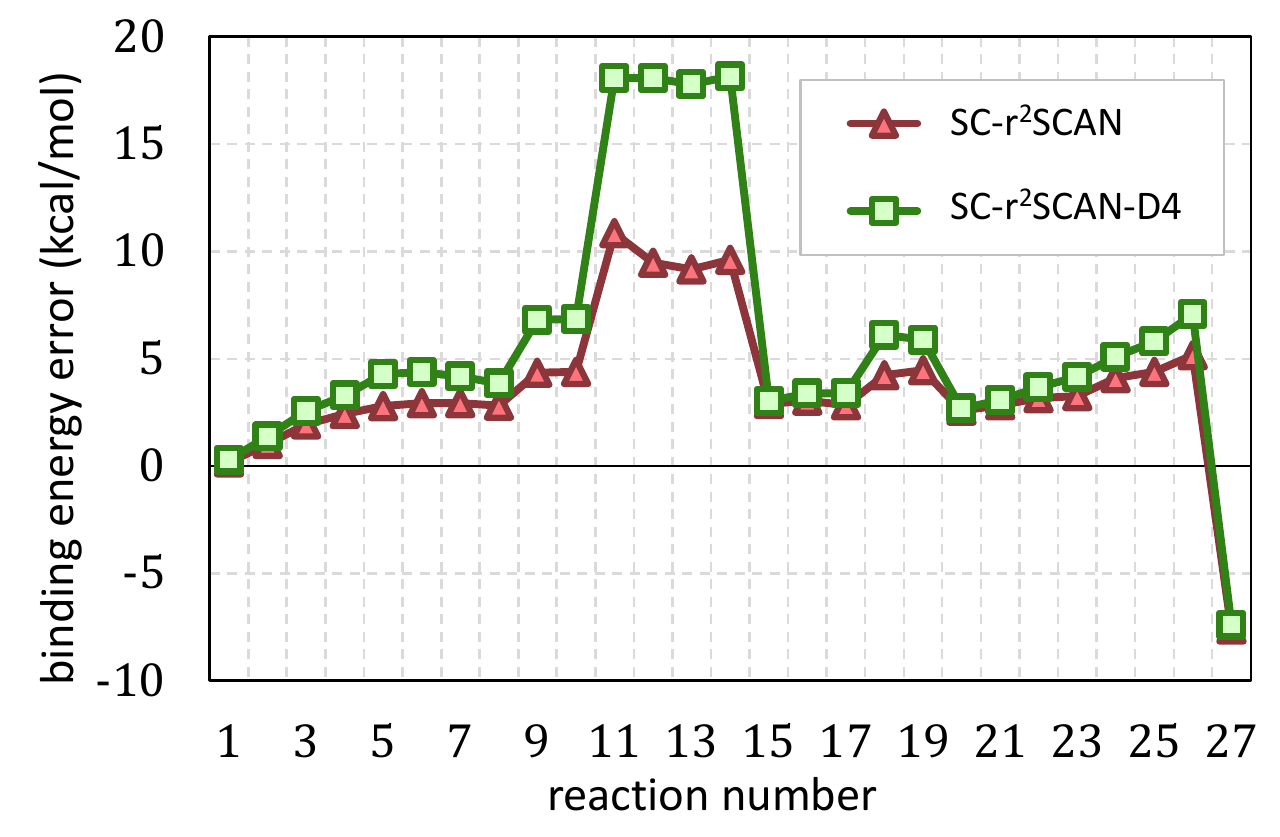}
\caption{
Binding energy error of WATER27 dataset
for SC-r$^2$SCAN and SC-r$^2$SCAN-D4.
The x-axis indicates the reaction number in WATER27
and the detailed information of reaction including geometries
can be found in the
GMTKN55 database.\cite{GHBE17}
D4 parameters of SC-r$^2$SCAN-D4 are from Ref.~\cite{EHNF21}.
}
\label{fgr:sc_water27}
\end{figure}

\begin{figure*}[htb]
\centering
\includegraphics[width=1.7\columnwidth]{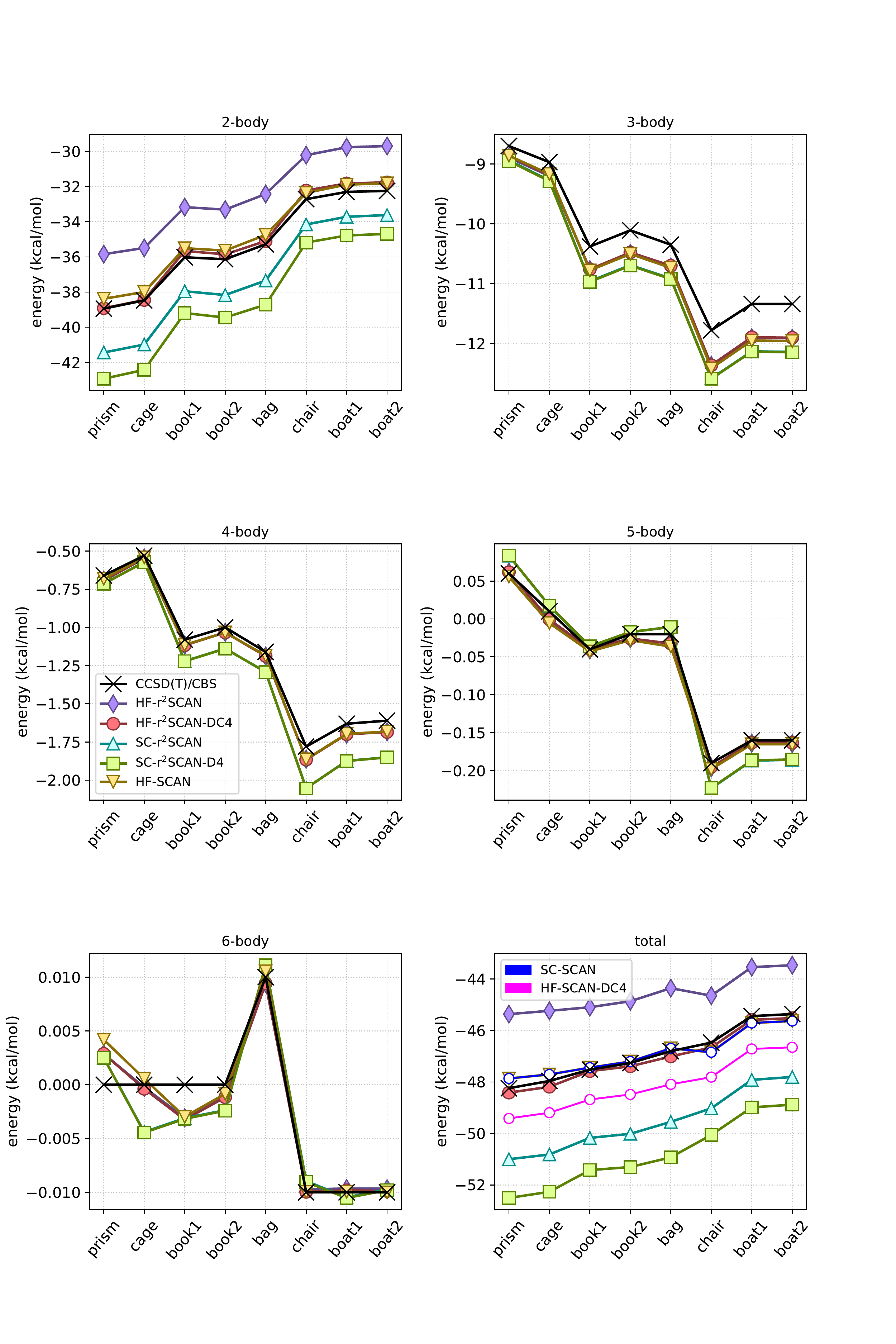}
\caption{
The  K-body energy plot corresponding to Figure~\ref{fgr:nbody}.
For comparison,
SC-SCAN and HF-SCAN-DC4 is additionally plotted in the total subplot.
def2qzvppd basis set is used.}
\label{fgr:nbodyerr}
\end{figure*}

\begin{figure}[htb]
\centering
\includegraphics[width=0.95\columnwidth]{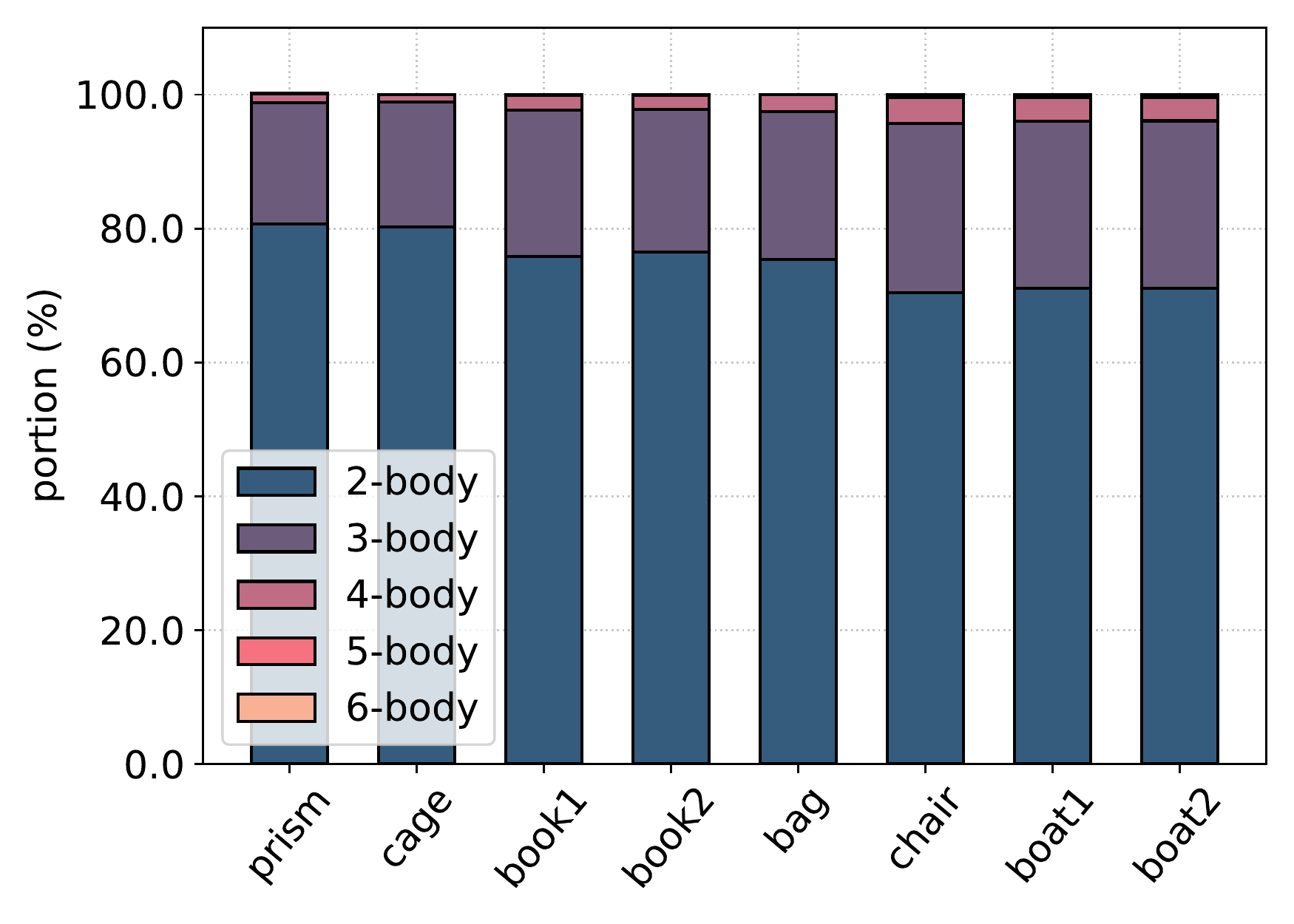}
\caption{
$K$-body percentage contributions to reference interaction energies of water hexamers (Figure~\ref{fgr:nbody}).
}
\label{fgr:nbodypor}
\end{figure}

\begin{figure}[htb]
\centering
\includegraphics[width=0.95\columnwidth]{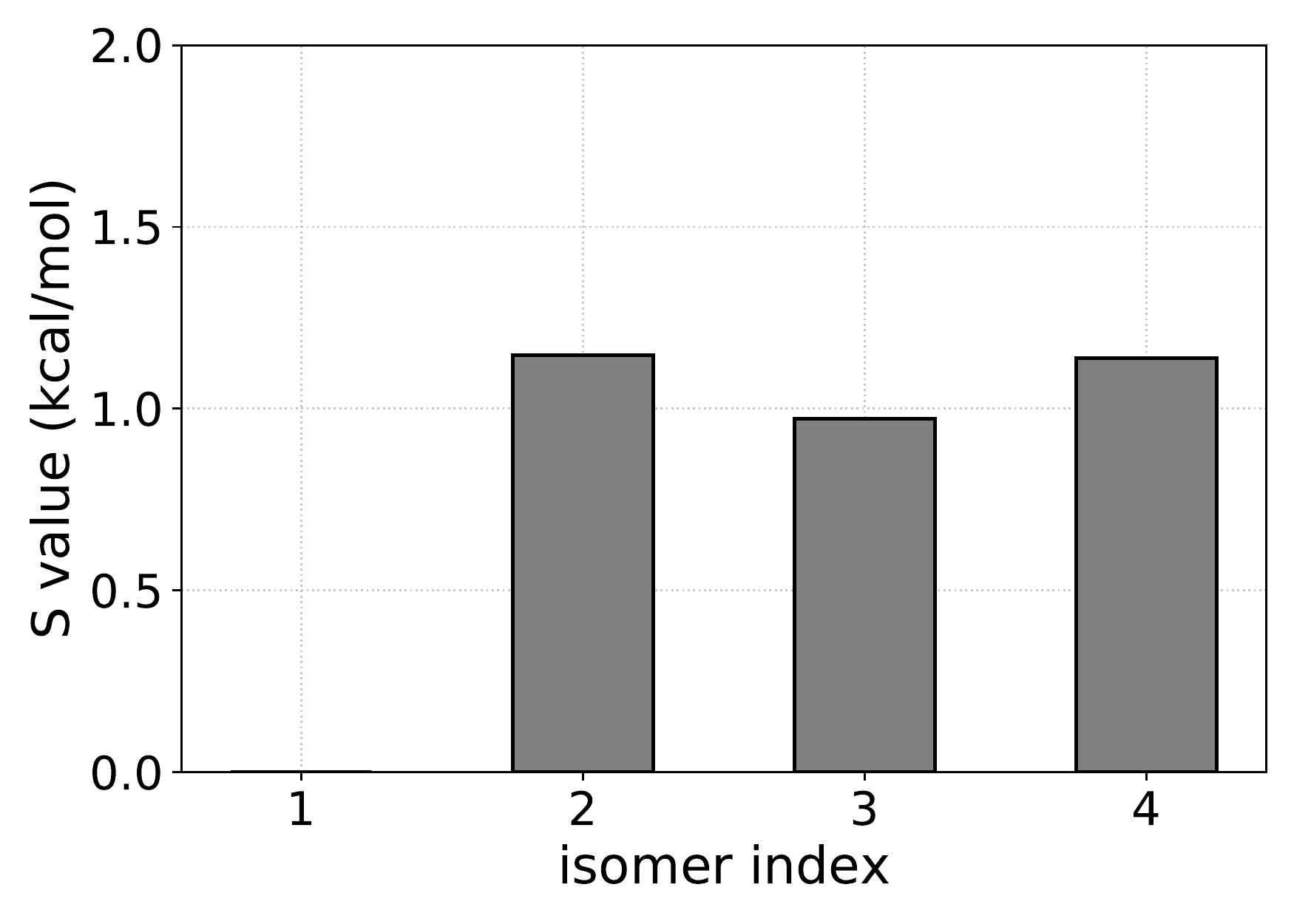}
\caption{
Density sensitivity $\tilde S$ value for 
isomer energies for the water 20-mers  corresponding to Figure~\ref{fgr:1}(d).
}
\label{fgr:20mer_s}
\end{figure}

\begin{figure}[htb]
\centering
\includegraphics[width=0.95\columnwidth]{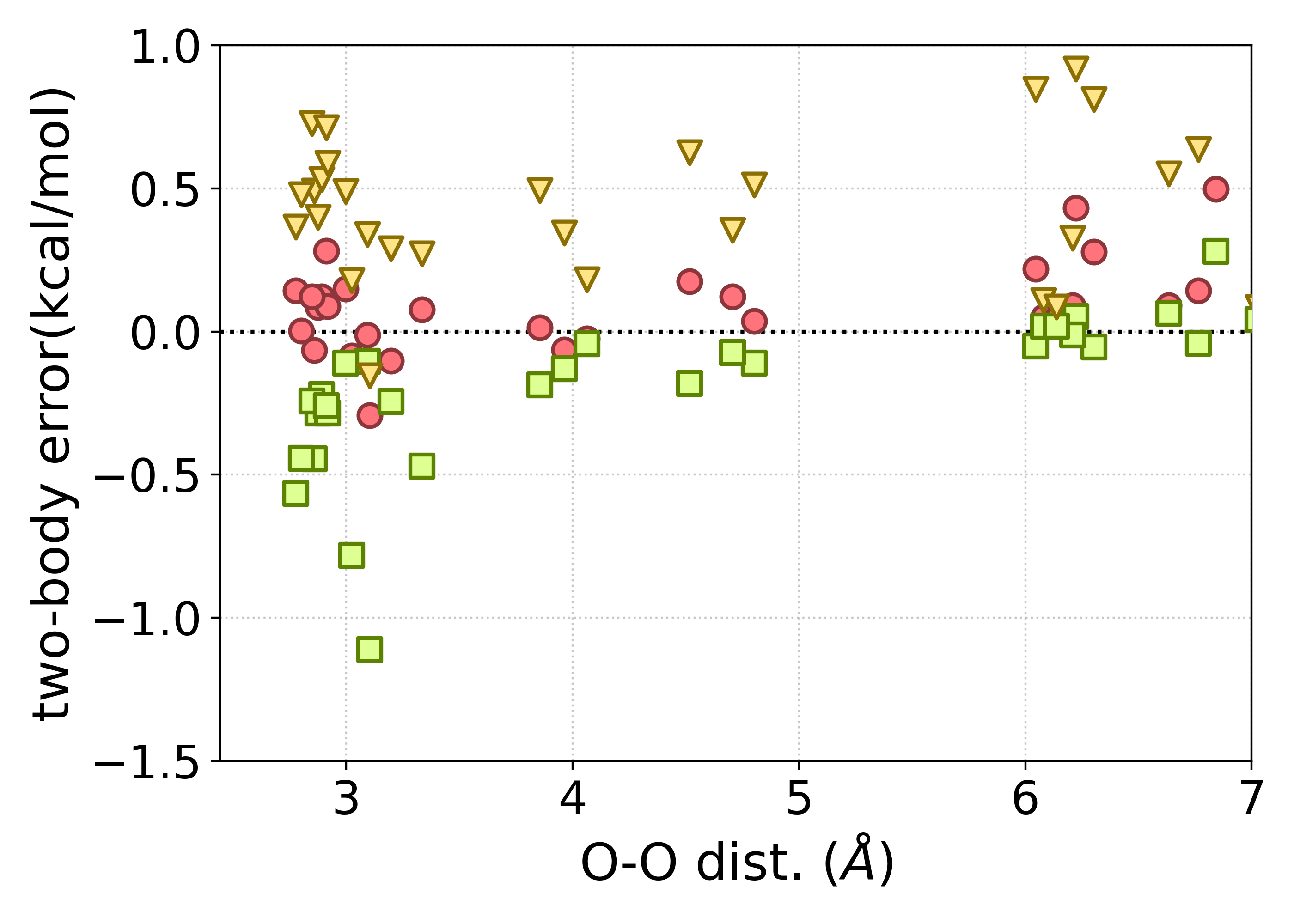}
\caption{
Aspirin$\cdots$water interaction including SC-r$^2$SCAN-D4.
The x-axis is the oxygen$\cdots$oxygen distance between
oxygen in the water and the specified oxygen in Aspirin (see inset of Fig~.\ref{fgr:1}(e)).
aug-cc-pvtz basis set is used.
}
\label{fgr:aspirin_tmp}
\end{figure}

\begin{figure*}[htb]
\centering
\includegraphics[width=1.95\columnwidth]{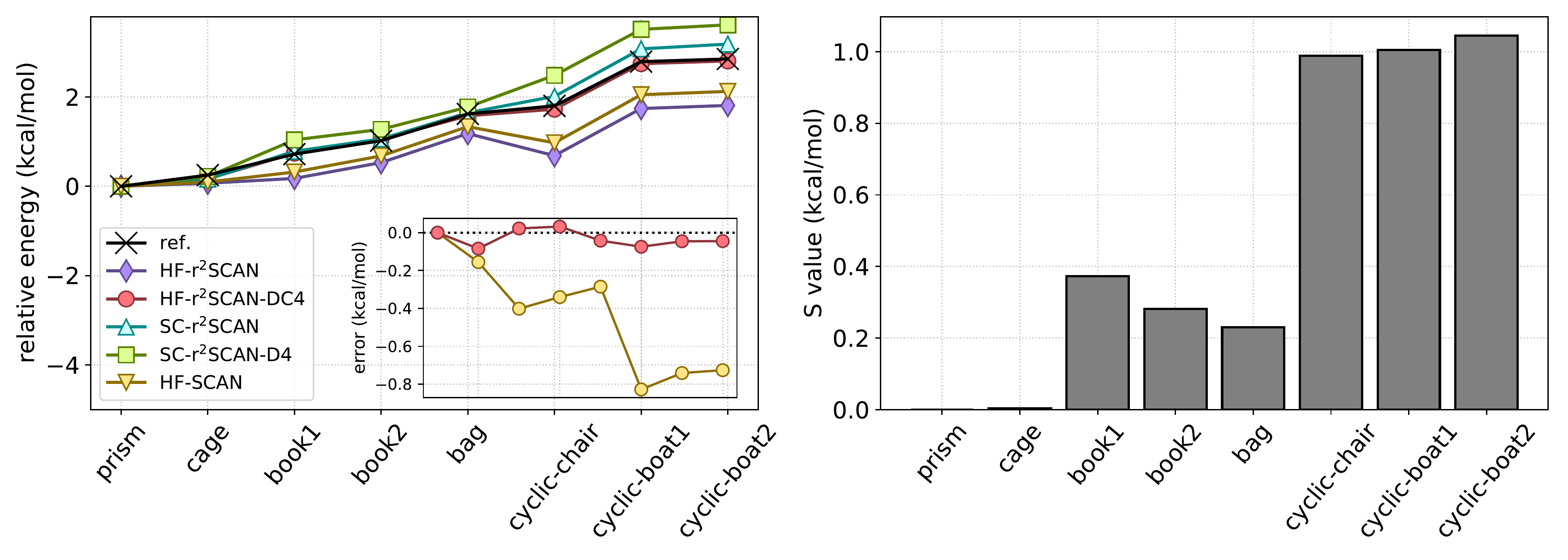}
\caption{
(left) Water hexamer isomer relative energy
compared to its global minimum geometry \textit{prism} structure;
(right) their $\tilde S$ value calculated from the r$^2$SCAN functional.
Geometries are from Ref.~\cite{RSBP16}.
MAEs of each functional are
0.61, 0.04, 0.13, 0.36, and 0.43 kcal/mol compared to the CCSD(T)/CBS from Ref.~\cite{OJ13}.
The ordering is the same as the legend ordering.
For HF-r$^2$SCAN-DC4,  we used the dispersion parameters
from Section~\ref{sec:d4para},  whilst
 for SC-r$^2$SCAN-D4,  we took the D4 
parameters from Ref.~\cite{EHNF21}.
The aug-cc-pvqz basis set was used for the calculations.
}
\label{fgr:hex_rank}
\end{figure*}

\begin{figure}[htb]
\includegraphics[width=0.95\columnwidth]{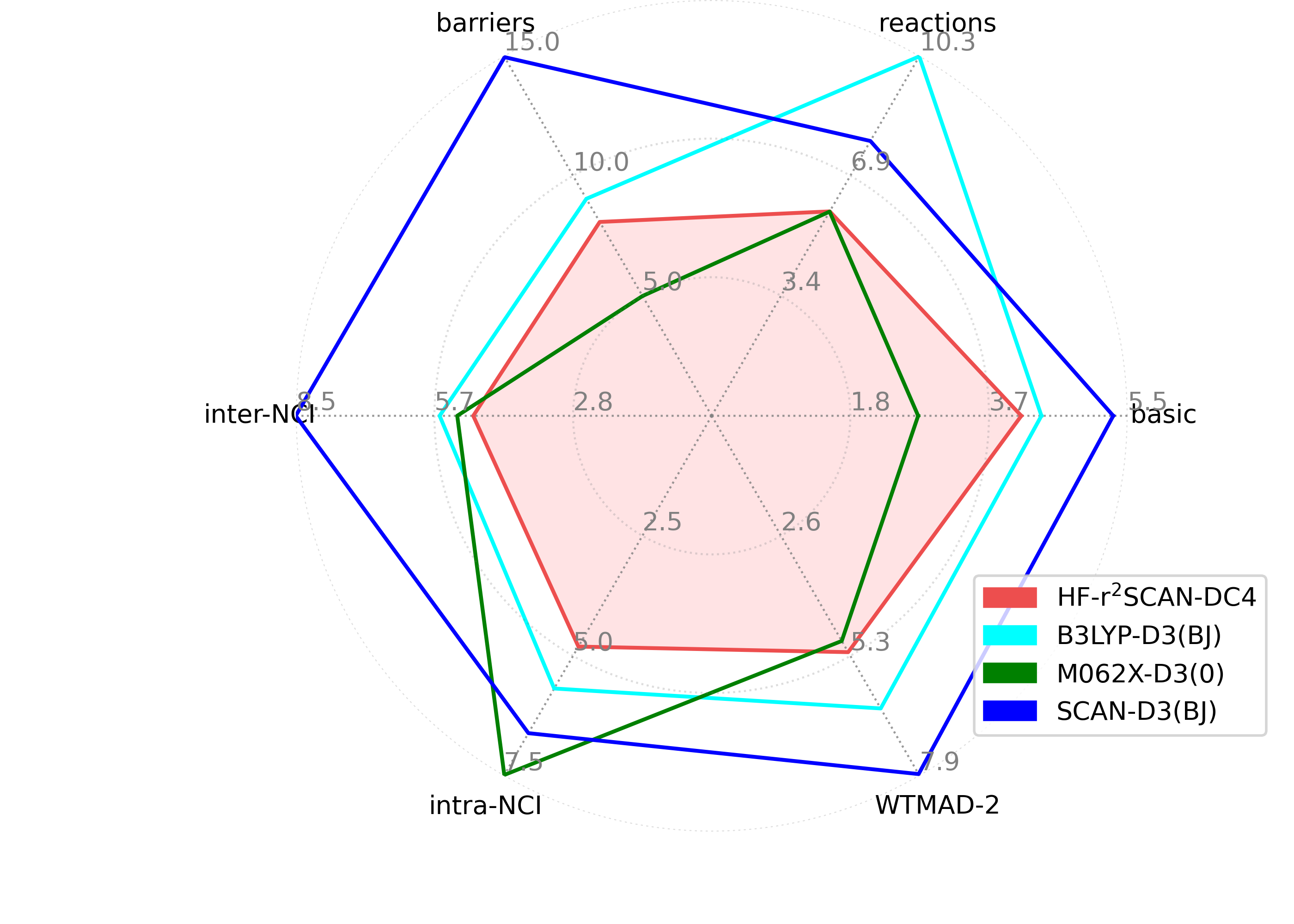}
\caption{
The hexagon plot with WTMAD-2 (kcal/mol) for all GMTKN55 and its categories for selected functionals.
(See Ref.~\cite{GHBE17} for the detailed description of the categories).
}
\label{fgr:g}
\end{figure}

\clearpage


\end{document}